\documentclass[a4paper,11pt]{article}
\pdfoutput=1 
\usepackage{jcappub} 
\usepackage[T1]{fontenc} 
\usepackage{booktabs}  
\usepackage{graphicx}  
\usepackage{subcaption} 
\usepackage{amsmath, bm}
\usepackage{mathtools}
\graphicspath{{./images/}}
\usepackage{comment}
\usepackage{xcolor} 

\newcommand{\HI}{\rm H{\sc i}}
\newcommand{\HII}{\rm H{\sc ii}}
\newcommand{\TB}{\delta T_{\rm b}}
\newcommand{\MSUN}{{\rm M}_{\odot}}
\newcommand{\XHI}{x_{\rm HI}}

\newcommand{\TS}{T_{\rm S}}
\newcommand{\TK}{T_{\rm K}}
\newcommand{\TCMB}{T_{\gamma}}

\newcommand{\OmegaB}{\Omega_{\rm B}}
\newcommand{\Omegam}{\Omega_{\rm m}}
\newcommand{\AVXHII}{\overline{x}_{\mathrm{HII}}}

\title{\boldmath Enhancing the detectability of ionized Regions during the Epoch of Reionization}

\author[a]{Rutvik Ashish Mahajan,}
\author[a, 1]{Raghunath Ghara,\note{Corresponding author.}}
\author[a]{Nishant Pradeep Deo,}
\author[b]{and Arnab Mishra}

\affiliation[a]{Department of Physical Sciences, Indian Institute of Science Education and Research Kolkata, Campus Road, Mohanpur, West Bengal-741246, India }
\affiliation[b]{Relativity \& Cosmology Research Centre, Department of Physics, Jadavpur University,
Kolkata 700032, India.}

\emailAdd{rutvikmahajan2003@gmail.com}
\emailAdd{ghara.raghunath@gmail.com}

\abstract{We present an improved matched filter method for detecting large ionized regions in 21 cm observations of the Epoch of Reionization. In addition to detection, the method constrains the properties of these regions, offering insights into the underlying source populations. Extending a previously developed Bayesian framework, we replace the spherical filter with an eight-parameter spheroidal filter, enabling a more flexible characterization of ionized bubbles. This enhancement significantly improves both detectability and recovery of bubble orientations. For a representative reionization scenario with mean ionization fraction $0.4$ at $z=7$, we find that a $10\sigma$ detection of the largest ionized region can be achieved with $\sim 1$ h of observations using the SKA-low AA4 and AA$^{\star}$ layouts. Our method can help identify regions in the observed field that host large ionized bubbles, making them prime targets for deeper follow-up observations.}

\begin{document}
\maketitle
\flushbottom

\section{Introduction}
\label{sec:intro}
The emergence of the first sources of radiation at the end of the Cosmic Dark Ages changed the Universe's thermal and ionization states from a cold and neutral phase to a hot and highly ionized one. The era when the primordial neutral hydrogen (\HI) in the intergalactic medium (IGM) became ionized is known as the Epoch of Reionization (EoR). The study of the mechanisms behind the formation of the first sources during the EoR, understanding their properties, and evolution is a topic of modern astronomy. Signals from these high-redshift sources are faint, and hence, using direct observation techniques with our currently available sources is extremely challenging. Nevertheless, various surveys have made significant progress towards detecting many high-redshift galaxies  \cite{Ellis13, Bouwens23, Harikane24} and quasars \cite{Fan06b, Mortlock11, Venemans15,  2018Natur.553..473B} over the last few decades.

An indirect way of detecting the high-redshift sources would be by looking at their effect on the \HI\ gas in the IGM. Radiation from early sources can modify the population of \HI\ in its hyperfine states, therefore impacting the 21 cm signal emission of \HI. Thus, the observation of the cosmological \HI\ signal from the EoR is an efficient way to detect the early radiating sources. Several radio observations have been set up to measure the EoR signal strength. Radio observations with large interferometers such as the Low Frequency Array \cite[LOFAR\footnote{\url{http://www.lofar.org/}};][]{2020MNRAS.493.1662M}, the Giant Metrewave Radio Telescope \cite[GMRT;][]{paciga13}, the Precision Array for Probing the Epoch of Reionization \cite[PAPER\footnote{\url{http://eor.berkeley.edu/}};][]{2019ApJ...883..133K}, the Murchison Widefield Array \cite[MWA\footnote{\url{https://www.mwatelescope.org/}};][]{2020MNRAS.493.4711T}, and the Hydrogen Epoch of Reionization Array \cite[HERA\footnote{\url{https://reionization.org/}};][]{Abdurashidova_2023} have attempted to measure the spatial fluctuations of the \HI\ signal from the EoR. These observations aim to measure the fluctuations in the EoR \HI\ signal in a statistical sense.  The upcoming radio interferometer, the Square Kilometre Array (SKA)\footnote{\url{http://www.skatelescope.org/}}, will have more sensitivity compared to the existing interferometers, and will also produce tomographic images of the signal \cite{2015aska.confE..10M, Koopmans_2015, ghara16}.

The expected \HI\ signal from the EoR IGM is significantly weaker than the galactic and extra-galactic foregrounds and the system noise. Recovering the faint signal from the radio interferometric measurement needs long observation hours and precise calibration during the data analysis phase to reduce artefacts caused by bright foreground sources 
\cite{2016MNRAS.461.3135B, 2024MNRAS.527.3517M}. Recent advancements in calibration techniques 
\cite[see, e.g.,][]{yatawatta20112011, 2019ApJ...884..105K, 2020ApJ...888...70K, 2020Mevius, Gan2022, 2023A&A...669A..20G, emilio2024}, foreground mitigation algorithms 
\cite[e.g.,][]{2014PhRvD..90b3019L, 2018MNRAS.478.3640M, 2024MNRAS.527.3517M, 2024MNRAS.527.7835A, 2025ApJ...988...84G}, ionospheric impacts 
\cite[e.g.,][]{2016RaSc...51..927M, 2021A&A...652A..37E} and mitigation of radio frequency interference (RFI)
\cite[see e.g.,][]{offringa2019, Wilensky_2019,  2025A&A...697A.203M}, along with enhanced sky-model construction 
\cite[see e.g.,][]{Patil2016, Ewall2017} during the data analysis process, are yielding encouraging outcomes in diminishing systematic effects, thereby aiding in the recovery of the targeted 21 cm signal.

The 21 cm signal is often characterized using statistical measures such as the power spectrum \cite{2015MNRAS.449.4246G, 2017MNRAS.468.3869S, 2018MNRAS.475.1213S, 2018MNRAS.475..438R}, the bispectrum \cite{2019JCAP...02..058G, 2022JCAP...11..001K, 2024JCAP...10..003N}, the Multi-frequency Angular Power Spectrum \cite{2023MNRAS.522.2188S} and tomographic analyses \cite{2018MNRAS.479.5596G, 2021JCAP...05..026K, 2024MNRAS.530..191G}. Its detectability has also been extensively explored using theoretical models. At present, various radio interferometer-based EoR observations have produced upper limits on the signal power spectrum. For example, upper limits from GMRT \cite{paciga13}, PAPER \cite{2018ApJ...868...26C, 2019ApJ...883..133K}, LOFAR  \cite{2019MNRAS.488.4271G, 2020MNRAS.493.1662M, 2024MNRAS.534L..30A, 2025arXiv250418534C, lofar2025}, MWA \cite{2019ApJ...884....1B, 2020MNRAS.493.4711T}, HERA \cite{Abdurashidova_2023} 
and the NenuFAR \cite{munshi2024, 2025MNRAS.tmp.1333M} are some of such results. These power spectra measurements have been used to constrain the properties of the EoR sources as well as the states of the EoR IGM \cite[see e.g.,][]{2020MNRAS.493.4728G, 2020arXiv200603203G, 2020MNRAS.498.4178M, 2022ApJ...924...51A, Ghara_2024, 2025A&A...699A.109G}.

A complementary method to efficiently detect the signature of the large structures of the \HI\ signal, such as the large ionized regions during the EoR, is by using the matched-filtering technique on the measured visibilities \cite{kanan2007MNRAS.382..809D}. The method enhances the detectability of \HII\ regions by using predefined filters on the recorded visibilities. Previous studies, such as \cite{kanan2007MNRAS.382..809D, 2012MNRAS.426.3178M, mishra2024} used real-space spherical top-hat filters to predict the detectability of the EoR \HII\ regions for different interferometers. Furthermore, the analysis in \cite{mishra2024} showed that this technique could also be used to measure the bubble radius and constrain the neutral hydrogen fraction in the surrounding IGM. An extension of this method to Cosmic Dawn has been used to predict detectability of isolated absorption or emission regions \cite[see e.g.,][]{ghara15c}.

Use of the image space spherical top-hat filter in the matched filtering technique is often motivated by the theoretical predictions of large spherical-shaped \HII\ regions around luminous quasars during the EoR.  The detection of luminous quasars at redshift, $z \gtrsim 7$ \cite{Mortlock11, 2018Natur.553..473B} suggests that the ionized regions around these sources will be significantly larger than those around galaxies that host only stars. Also, the locations of these quasars can be used for targeted \HII\ region searches using matched filtering techniques \cite{datta2012a}, and also for constraining the properties of the quasars \cite[see e.g.][]{2012MNRAS.426.3178M}.

In general, such favourable situations of having known bright quasars in the interferometric field of observation with ideal foreground properties may not be common. In a blind search of the largest \HII\ regions in the field of view, where no prior information about the locations of the ionized regions is known, one needs to vary the location of the filters in addition to other parameters such as the radius of the top-hat filter. Such a scenario was considered in  \cite{2020MNRAS.496..739G}. The study used spherical top-hat filters in image space and showed that even $\approx 20$ h of integration with the AA4 layout of SKA-low will be able to detect and constrain the properties of an \HII\ region of size $\gtrsim 50$ Mpc at $z\simeq 7$.

The morphology of the \HII\ regions during the EoR is expected to be complex after the overlap between individual ionized regions started \cite{Furlanetto_2004,McQuinn_2007}. As galaxy formations occur mostly on the cosmic filaments, the ionized regions are likely to overlap and create a tunnel type structure. This can happen as early as the Universe had a mean ionization fraction $\AVXHII \gtrsim 0.1$ \cite[see e.g.][]{2024MNRAS.530..191G}. At a later time, more overlap can create a more complex morphology of the \HI\ distribution in the IGM. In general, the shape of the largest ionized region could be non-spherical. Use of a real-space top-hat filter in the matched filtering technique, therefore, might not be an optimum choice.

In this work, we extend the matched filtering method adopted in \cite{2020MNRAS.496..739G} and use a spheroidal filter to enhance the detectability as well as extract more information about the EoR ionized regions. We apply this Bayesian framework to simulated mock visibilities containing the expected signal and system noise of different configurations of SKA-low, e.g., AA2, AA$^{\star}$, and AA4. This paper is structured as follows. In Section \ref{sec:methodology}, we discuss the framework of our spheroidal filter. We present our results in Section \ref{sec:res} before we conclude in Section \ref{sec:con}. This study is based on $\Lambda$CDM cosmological model with cosmological parameters $\Omegam=0.32$, $\Omega_\Lambda=0.68$, $\OmegaB=0.049$, $h=0.67$, $n_{\rm s}=0.96$, and $\sigma_8=0.83$, which are consistent with $Planck$ measurements \cite{Planck2013}.


\section{Methodology}
\label{sec:methodology}
In this section, we describe the different layouts of SKA-low considered in this study, the method to generate the mock visibilities and the basics of the matched filtering techniques, along with the parameters of the spheroidal filter.

\subsection{Mock SKA-low observations}
\label{sec:obs}
\begin{figure}[]
\centering 
\includegraphics[width=\textwidth]{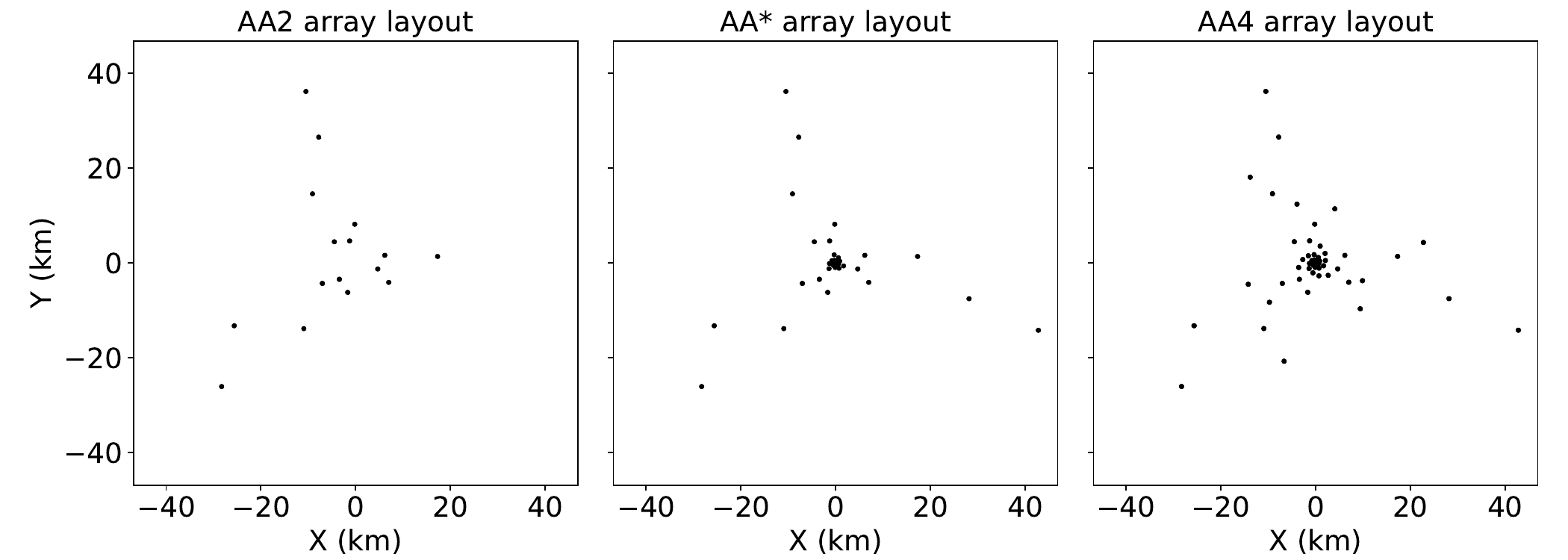}
\caption{\label{fig:AAs} Layouts of the three stages in the construction of the SKA-low telescope. The panels from left to right represent AA2, AA$^{\star}$, and AA4 SKA-low configurations with $68$, $307$, and $512$ antennas, respectively.}
\end{figure}

Before reaching its full form AA4 with $512$ antennas, we will have observations with two intermediate SKA-low configurations, namely the AA2 and AA$^{\star}$ with $68$ and $307$ antennas, respectively. We consider these three configurations while choosing the AA4 as our fiducial SKA-low layout. Figure \ref{fig:AAs} displays the layouts of the antennas in each of the three configurations\footnote{\url{https://gitlab.com/ska-telescope/ost/ska-ost-array-config/-/blob/master/docs/examples.ipynb)}}. We note that the layout of AA4 is slightly different than the one used in \cite{2020MNRAS.496..739G} while the number of antennas remains the same.

The quantity that is measured from a radio observation is the visibility, $V(\vec{U}, \nu)$. This quantity depends on the baseline vector $\vec{U}$ and the frequency of observation $\nu$. The baseline vector is defined as \(\vec{U} = \vec{d}/\lambda\), where \(\vec{d}\) is the distance vector between two antennas, and $\lambda$ is the wavelength of observation. 

\begin{figure}[]
\centering 
\includegraphics[width=\textwidth]{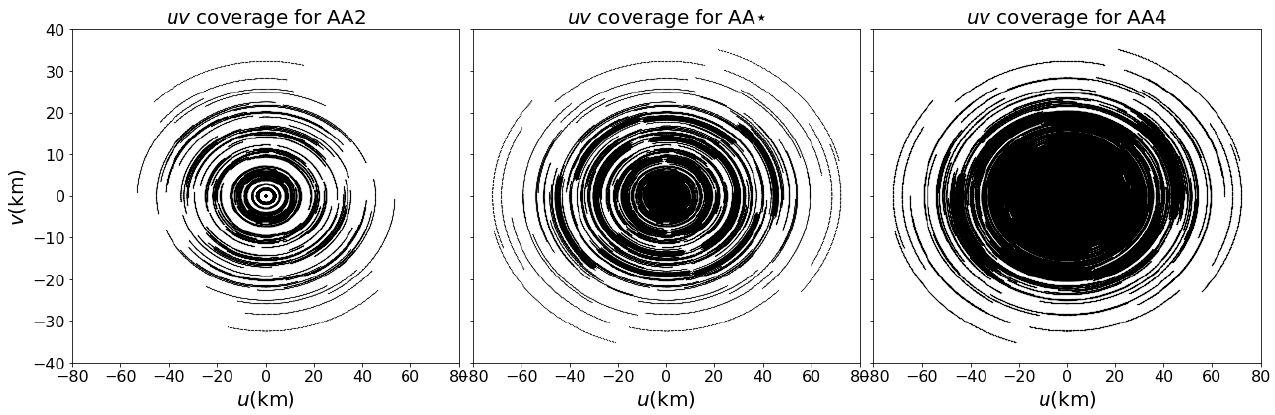}
\caption{\label{fig:AAs_UV_coverage} UV coverage at frequency $\nu=177$ MHz for the three SKA-low array layouts at different construction stages. The panels from left to right represent AA2, AA$^{\star}$, and AA4 SKA-low configurations, respectively.  These correspond to mock observations at a part of the sky with right ascension $0^{\circ}$ and declination $-30^{\circ}$ for $4$ h daily. }
\end{figure}

\begin{figure}[]
\centering 
\includegraphics[width=0.7\textwidth]{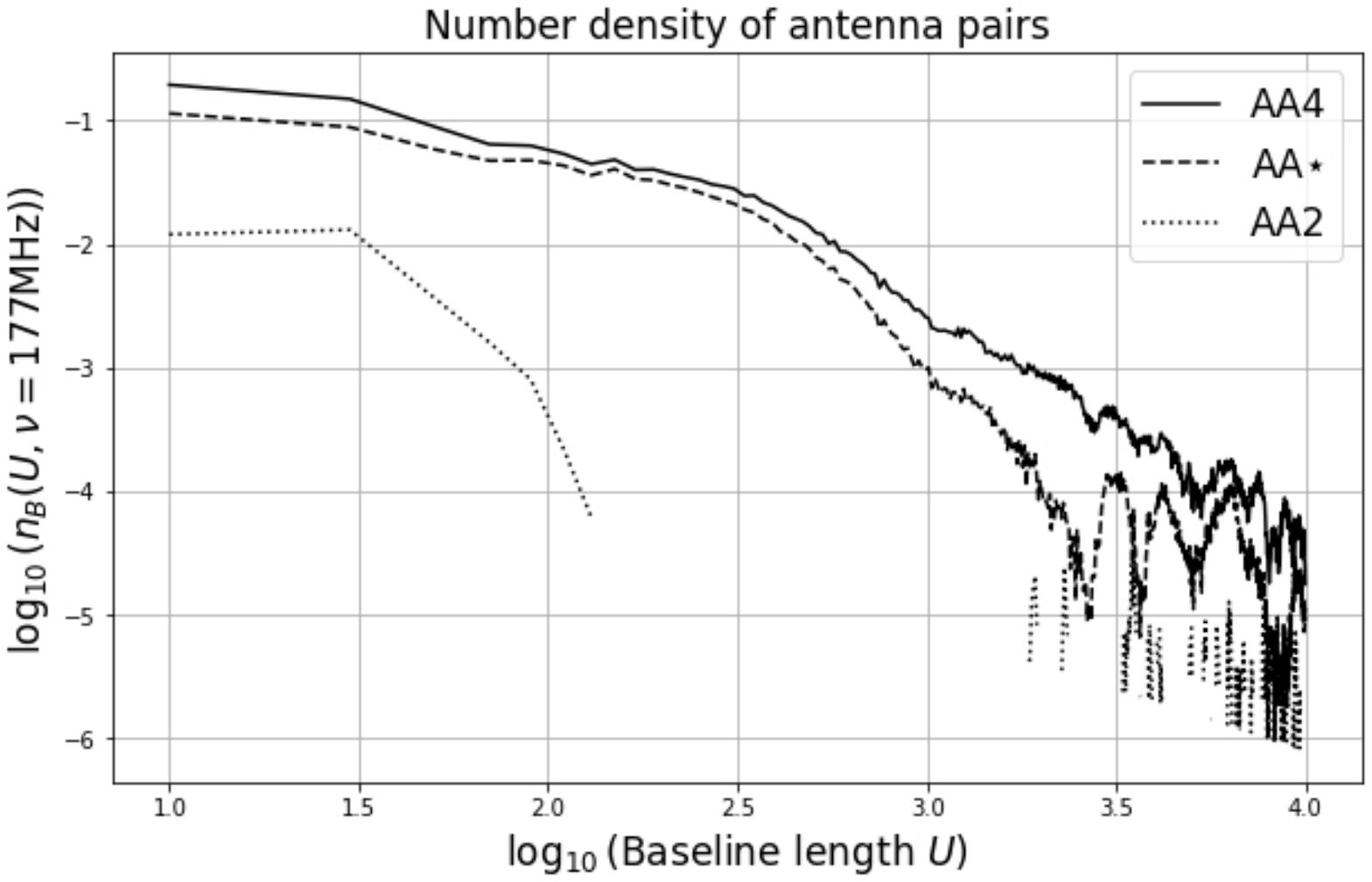}
\caption{\label{fig:baseline}This displays the relation of the number density of antenna pairs having baseline $U$ at frequency 177 MHz for different configurations of SKA-low.}
\end{figure}

Table \ref{tab:obs_params} shows the details of the SKA-low mock observations, which are done at a part of the sky with right ascension $0^{\circ}$ and declination $-30^{\circ}$ for $4$ h daily. The observation details are the same as considered in \cite{2020MNRAS.496..739G}. For such observations, the $uv$ coverages of these SKA-low configurations are shown in Figure \ref{fig:AAs_UV_coverage}. The effect of less number of antennas in the AA2 configuration is prominently visible in the AA2 $uv$ coverage as shown in the left panel of Figure \ref{fig:AAs_UV_coverage}. The same is also visible in Figure \ref{fig:baseline}, which displays $n_B(U,\nu)$, the number density of antenna pairs having baseline $U$ at frequency $\nu=177$ MHz, against the baseline $U$. Here, $\int d^2U n_B(U,\nu)= N_{\rm ant}\times(N_{\rm ant}-1)/2$, where $N_{\rm ant}$ is the number of antennas in the radio interferometer. For all three configurations, we see that the baseline density decreases as the baseline length increases.

\begin{table}[h]
    \centering
    \renewcommand{\arraystretch}{1.3} 
    \begin{tabular}{l c}  
        \toprule
        \textbf{Parameter} & \textbf{Value} \\ 
        \midrule
         Number of antennas ($N_{\text{ant}}$) & 512, 307, 68 (AA4, AA$^{\star}$, AA2) \\
        Effective area ($A_{\text{eff}}$) & 962 m$^2$ \\
        Integration time ($\Delta t_c$) &  120 sec\\
        Observation time per day & 4 h \\
        Total observation time ($t_{\text{obs}}$) & 20 h (AA4), 55 h (AA$^{\star}$), 1100 h (AA2) \\
        Declination to observed region & -30$^\circ$ \\
        Right Ascension & 0$^\circ$\\
        Observation frequency ($\nu_c$) & 177.5 MHz\\
        Bandwidth ($B_{\nu}$) &  18.94 MHz\\
        Frequency resolution ($\Delta\nu$) & 175 kHz \\
        System temperature ($T_{\text{sys}}$) & 
        $\left[ 100 + 60 \times \left( \frac{\nu}{300\,\text{MHz}} \right)^{-2.55} \right]$ K \\
        \bottomrule
    \end{tabular}
    \caption{The details of the mock observations with SKA-low. }
    \label{tab:obs_params}
\end{table}

In real observations, the measured visibilities contain not only the 21 cm signal that we are interested in, but also the foregrounds from astrophysical sources, instrumental noise, and other unwanted signals such as RFI. Such observations also face calibration issues. Here, we consider a simplified scenario, which assumes that foregrounds and other systematic effects can be perfectly modelled and removed. Following those steps, what will remain in the visibility are the 21 cm signal $S(\vec{U},\nu)$ and the noise $N(\vec{U},\nu)$. 
\begin{equation}
    V(\vec{U},\nu) = S(\vec{U},\nu) + N(\vec{U},\nu).
\end{equation}
In the following sub-sections, we briefly describe the methodology of simulating $S(\vec{U},\nu)$ and  $N(\vec{U},\nu)$, which are later used in this study.

\subsubsection{Simulating signal $S(\vec{U},\nu)$}
\label{sec:simdtb}
Here we use the same simulated differential brightness temperature ($\TB$) fields of the 21 cm signal as earlier used in \cite{2020MNRAS.496..739G}. We note that, in addition to the $\TB$ field, $S(\vec{U},\nu)$ also depends on the baseline distributions of the SKA-low configurations. Here we briefly describe the steps adopted to obtain $S(\vec{U},\nu)$. We refer the reader to \cite{2020MNRAS.496..739G} for the detailed procedure of obtaining the simulated $S(\vec{U},\nu)$.

First, a dark matter-only $N$-body simulation was carried out using the code {\sc cubep$^3$m} \cite{Harnois12} within a simulation box of size 200 $h^{-1}$ comoving Mpc. This provided us with snapshots of density and velocity fields as well as list of dark-matter halos from $z$ range from 20 to 6, with a time difference between two successive snapshots of 10 Myr. The lists of dark matter halos contain halos with mass larger than $2.2\times 10^9 ~\MSUN$. The same $N$-body simulation was previously used in \cite[e.g.][]{ghara15b, 2019MNRAS.487.2785I, 2020MNRAS.496..739G}. 

The redshift-evolution of the gas temperature ($\TK$), spin temperature ($\TS$) and the neutral fraction ($\XHI$) in the simulation box was performed using {\sc grizzly} code \cite{ghara15a}, which is based on a one-dimensional radiative transfer scheme. The coeval simulation was performed on a $216^3$ grid. This simulation assumed that each dark matter halo contains a galaxy that emit UV as well as X-ray photons. The rate of emission of ionizing photons per second per stellar mass is considered as $\approx 5.67\times 10^{43}$. The X-ray luminosity is taken as 5$\%$ of the UV luminosity, while the spectral index of the X-ray spectral energy distribution is taken as 1.5. We name this scenario as \textit{Galaxy-noQSO}. In addition, we consider two more reionization scenarios where we assume that the most massive halo within the simulation box at redshift 7 contains a Quasar with an age of 10 Myr and 30 Myr\footnote{Previous studies using quasar proximity zones (e.g., \cite{morey2021estimating, eilers2021detecting}) suggest that quasar lifetimes during the EoR may be shorter than the values we assume here}, respectively. We named them as \textit{Galaxy-QSO10Myr} and \textit{Galaxy-QSO30Myr}, respectively.  We follow the choices as described and adopted by \cite{Mortlock11, 2012MNRAS.426.3178M}, and adjust this quasar's luminosity such that it emits $1.3 \times 10^{54}$ ionizing photons per second.

Next, we generate the coeval fields of the brightness temperature from the density contrast ($\delta_{\rm B}$), $\XHI$ and $\TS$ using \cite[see e.g.,][]{madau1997, Furlanetto2006}
\begin{align}
 \TB (\bm{x}, z) = 27 ~ x_{\rm HI} (\bm{x}, z) [1+\delta_{\rm B}(\bm{x}, z)] \left(\frac{\OmegaB h^2}{0.023}\right) 
\times \left(\frac{0.15}{\Omegam h^2}\frac{1+z}{10}\right)^{1/2}\left[1-\frac{\TCMB(z)}{T_{\rm S}(\bm{x}, z)}\right]\,\rm{mK}.
\label{brightnessT}
\end{align}
Here,  $\bm{x}$ is the three-dimensional position in the simulation box.  $\TCMB(z)$ = 2.73 $\times (1+z)$ K is the Cosmic microwave background (CMB) temperature at $z$. Next, we include the redshift space distortions effect to the $\TB$ cubes using the line of sight velocity fields using the Mesh-to-Mesh Real-to-Redshift-Space-Mapping scheme \cite{mao12, ghara15b, 2021MNRAS.506.3717R}.  In addition, we account for the evolution of the signal with redshift or so-called \textit{Light-cone} effect \cite{ghara15b}. Finally, we reduce the resolution of the $\TB$ light-cone by a factor of 2 to reduce the computation time of our Bayesian analysis. This set the angular resolution of the map at $z\approx 7$ to $1.075'$, which corresponds to a comoving length scale of $\approx 2.76$ Mpc.

\begin{figure}
    \centering
  \includegraphics[width=\textwidth]{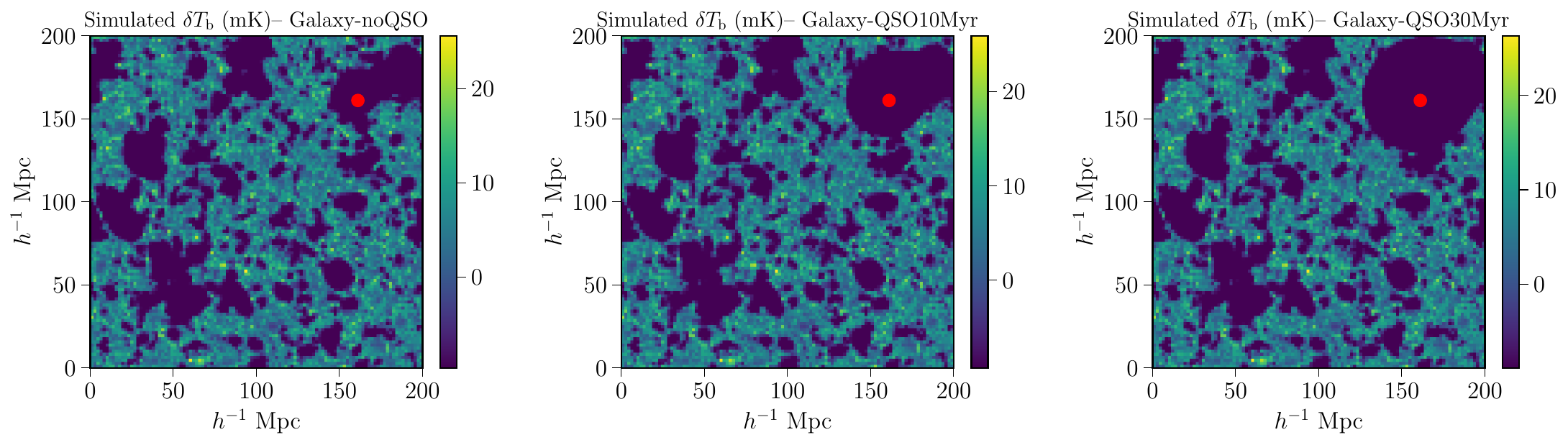}
    \caption{\label{fig:signal_all} Two-dimensional slices of the simulated $\TB$ light cones. The left-to-right panels correspond to considering the \textit{Galaxy-noQSO}, \textit{Galaxy-QSO10Myr} and \textit{Galaxy-QSO30Myr} scenarios, respectively, at $z=7$. The red dot points show the location of the rare quasar formed in the most massive dark matter halo in the simulation box.}
\end{figure}

Figure \ref{fig:signal_all} shows the comparison between the $\TB$ maps that contain the most massive dark matter halo at redshift 7. The volume-averaged ionization fraction at this stage is $\approx 0.4$.  From left to right, the panels represent the models \textit{Galaxy-noQSO}, \textit{Galaxy-QSO10Myr}, and \textit{Galaxy-QSO30Myr}, respectively. Clearly, the structure of the ionized region around the most massive dark matter halo is more complex in the absence of a bright quasar. However, the region around the most massive dark matter halo is still quite big in volume due to the presence of multiple nearby galaxies whose individual ionized regions have overlapped into the big ionized region. The Bayesian analysis on the \textit{Galaxy-noQSO} as performed in \cite{2020MNRAS.496..739G} identified the bubble around the most massive dark matter halo as the one with the largest SNR.

\begin{figure}
    \centering
  \includegraphics[width=\textwidth]{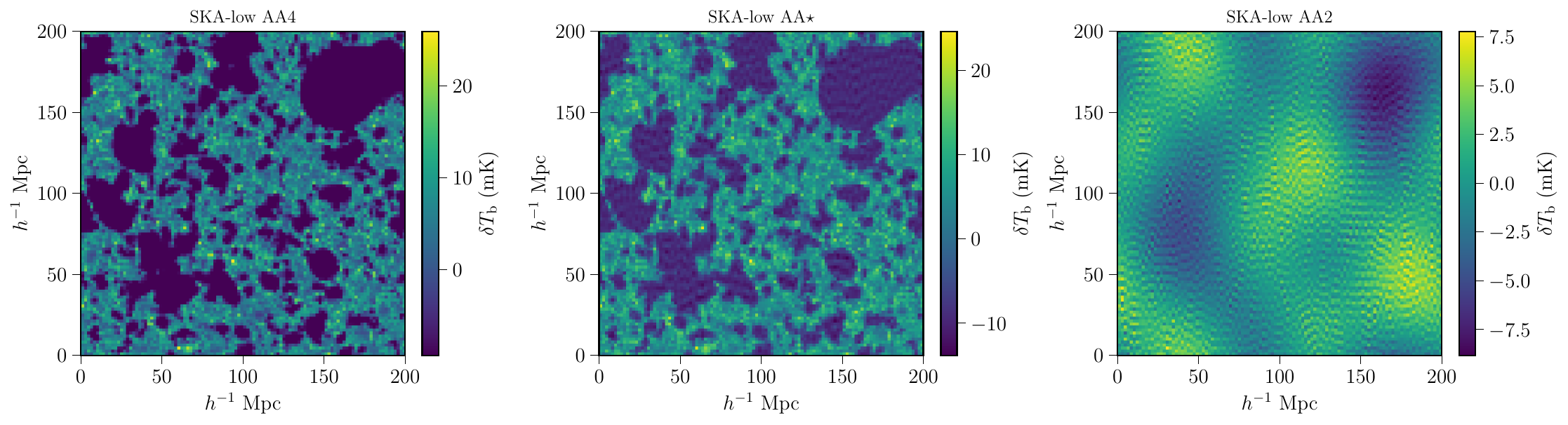}
    \caption{\label{fig:signal} Two-dimensional slices of the brightness temperature from the fiducial simulated light-cone considering the \textit{Galaxy-QSO10Myr} scenario and an observation time of 20 h, selected around the most massive dark matter halo. From left to right, the panels show the expected image (in the absence of noise) at $z=7$ using the SKA-low AA4, AA$^{\star}$, and AA2 configurations. The original simulated map resolution is $1.075$ arcmin, corresponding to a physical scale of $2.76$ Mpc. }
\end{figure}

Next, we produce the gridded signal visibilities $S(\vec{U},\nu)$ of $\TB (\bm{x}, z)$  by performing a discrete Fourier transform of the two-dimensional slice from the Light cone.
\begin{equation}
S(\vec{U}, \nu) = F(\vec{U}, \nu) \int d^2\theta ~I_{\rm S}(\vec{\theta}, \nu)~A(\vec{\theta})~e^{i2\pi \vec{\theta} \cdot \vec{U}},
\label{visibi}
\end{equation}
where the sky specific intensity at frequency $\nu$, $I_{\rm S}(\vec{\theta},\nu)$, can be written as,
\begin{equation}
I_{\rm S}(\vec{\theta},\nu) = \frac{2 k_{\rm B} \nu^2  }{c^2} \TB (\vec{\theta}, \nu).
\label{inten}
\end{equation}
Here we have done a conversion of three-dimensional position $\bm{x}$ in the Light-cone box to an angle $\vec{\theta}$ on the sky and a frequency $\nu$ along the line of sight. The quantity $A(\vec{\theta})$ in equation \ref{visibi} is the primary beam pattern of an individual SKA-low antenna. Our simulation box corresponds to a field of view of $1.94^{\circ}\times1.94^{\circ}$ at $z\approx7$, which is rather small compared to the SKA-low primary beam size. This allows us to set $A(\vec{\theta}) \approx 1$ throughout the study. The quantity $F(\vec{U}, \nu)$ in equation \ref{visibi} is the sampling function of the SKA-low $uv$ coverage, i.e., $F = 1$ for the grid points with at least one measured $V(\vec{U},\nu)$, and $F = 0$ otherwise.

Figure \ref{fig:signal} shows the expected images of $\TB$ extracted from the fiducial simulated light-cone for different SKA-low configurations. The same slice contains the centre of the most massive dark matter halo. Clearly, the unfilled $uv$ coverage in the AA$^{\star}$ and AA2 configurations distorted the $\TB$ maps. Additionally, the probed resolution for the AA2 case is much smaller than the resolution of the other two cases.

\subsubsection{Simulating noise $N(\vec{U},\nu)$}
\label{sec:sim_ns}
Assuming the system noise at different baselines and frequency channels are uncorrelated, the rms noise at each baseline and frequency channel for an observation time of $t_{\rm obs}$, can be expressed as \cite{2020MNRAS.496..739G},

\begin{equation} \label{eqn:noise-gaussianuv}
    \sigma_N = \frac{\sqrt{2}k_B T_{\text{sys}}}{A_{\text{eff}}\sqrt{\Delta\nu_c t_{\text{obs}}}}.
\end{equation}
Here, $\Delta \nu_c$ denotes the frequency channel width, $t_{\text{{obs}}}$ denotes the observation time, $A_{\text{eff}}$ denotes the effective collection area of each antenna, and $T_{\text{sys}}$ denotes the system temperature (see Table \ref{tab:obs_params}). The frequency resolution ($\Delta\nu_c$) is set to 175 kHz, which corresponds to the same length scale associated with the angular resolution $1.075'$. Here we consider the same method as adopted in \cite{2020MNRAS.496..739G} for generating $N(\vec{U},\nu)$. The method first generates a Gaussian random distribution with zero mean and rms as described in equation \ref{eqn:noise-gaussianuv}. Thereafter, we multiply $F(\vec{U},\nu)$ to obtain the noise visibilities $N(\vec{U},\nu)$.

The real space noise maps $N(\vec{\theta},\nu)$ can be generated by performing Fourier transforms of $N(\vec{U},\nu)$ maps. The rms of the real space noise maps can be expressed as 
\begin{equation} \label{eqn:noise-gaussianreal}
    \sigma_{\rm rms} = \frac{\sqrt{2}k_B T_{\text{sys}}}{A_{\text{eff}}\sqrt{B_\nu t_{\text{obs}}N_{\text{ant}}(N_{\text{ant}}-1)/2}}.
\end{equation}
The frequency band chosen in this study is $B_\nu=18.94$ MHz, which corresponds to the size of our simulation box at $z\approx7$. Here, $N_{\rm ant} (N_{\rm ant} - 1) / 2$ is the number of baselines of the interferometer with $N_{\rm ant}$ number of antennas. We note that both $N(\vec{U},\nu)$ and $N(\vec{\theta},\nu)$ differ with SKA-low antenna configuration due to the change of $F(\vec{U},\nu)$ and $N_{\rm ant}$.  

We chose a total observation time ($t_{\rm obs}$) of 20 h for AA4 SKA-low configuration. This is the same as taken in \cite{2020MNRAS.496..739G}. For the mock observations with the other two configurations, we estimate $t_{\rm obs}$ so that $\sigma_{\rm rms}$ remains the same. The observation times for the AA$^{\star}$ and AA2 configurations are 55 and 1100 h, respectively.

\begin{figure}
    \centering
  \includegraphics[width=\textwidth]{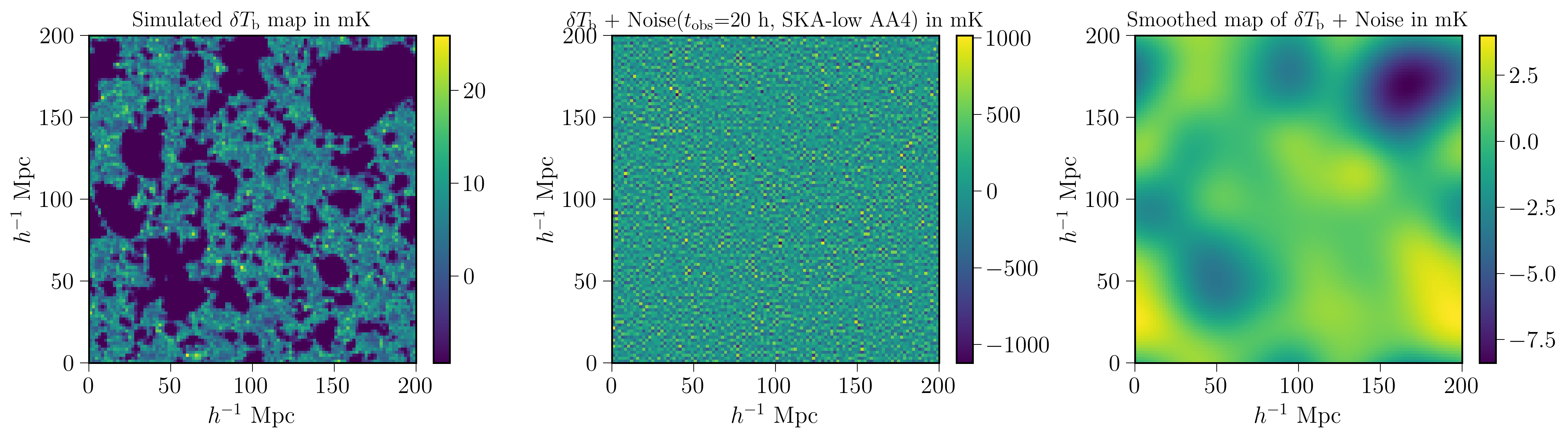}
    \caption{\label{fig:noise} Comparison between the brightness temperature slice from \textit{Galaxy-QSO10Myr} scenario and noise map for 20 h of observation at redshift 7 with SKA AA4. The left and the middle panels have resolutions of $1.075$ arcmin. The right panel shows the smoothed map of the middle panel where we have applied a Gaussian smoothing with $\sigma=9$.} 
\end{figure}

The left-to-right panels of Fig.~\ref{fig:noise} show (i) the simulated $\TB$ map for the \textit{Galaxy–QSO10Myr} case, (ii) the $\TB$ map combined with the SKA-AA4 noise realization for a 20-hour observation, and (iii) the smoothed version of the middle panel obtained using a Gaussian filter. In the noise-added map (middle panel), the ionized structures are almost entirely obscured, as the noise amplitude exceeds that of the underlying signal. When Gaussian smoothing is applied, the large-scale morphology of the ionized regions begins to re-emerge. However, because the smoothing process removes sharp boundaries, the identification of ionized regions in the smoothed map becomes sensitive to the chosen threshold.

Next, we estimate the maximum achievable SNR, defined as the ratio of the strongest signal to the rms noise in the image, by applying Gaussian filters with a range of kernel sizes. For SKA-AA4, a 20-hour observation at $z = 7$ yields a maximum SNR of approximately 13 after optimal smoothing. In contrast, the matched-filtering technique can achieve a similar SNR with a significantly shorter observing time (about 1 hour; see \cite{2020MNRAS.496..739G}). This demonstrates that matched filtering is a highly efficient method for indirectly detecting the 21-cm signal compared to simple image-domain smoothing.

\subsection{Bayesian framework of the matched filtering method}
\label{sec:bfmf}
Here, we describe the matched filter-based Bayesian framework to identify the largest ionized regions during the EoR efficiently. The main principle of this framework is to smooth $V(\vec{U},\nu)$ using an appropriate filter $S_f$ and define a likelihood $\Lambda$ which can be used for Bayesian analysis. 
For a uniform prior, the logarithm of the likelihood for a known signal $S_f(\vec{U}, \nu; \bm{\mu})$, characterized by the parameters $\bm{\mu}$, present in $V(\vec{U}, \nu)$ can be expressed as  \cite{1992PhRvD..46.5236F, 2020MNRAS.496..739G},
\begin{equation}
  \log \Lambda(\bm{\mu}) = \frac{1}{\sigma_{\rm rms}^2} \int {\rm d}^2 U \int {\rm d}\nu  ~\rho_B(\vec{U},\nu)
  \times \left[2 V(\vec{U},\nu) ~S_f^{*}(\vec{U}, \nu; \bm{\mu}) - |S_f(\vec{U},\nu; \bm{\mu})|^2\right],
\label{equ_exp_like}
\end{equation}
where, $\rho_B(\vec{U},\nu)$ is normalized baseline density, $\int {\rm d}^2 U \int {\rm d}\nu  ~\rho_B(\vec{U},\nu) = 1$. We use this $\log \Lambda(\bm{\mu})$ in a Bayesian framework, namely the Markov Chain Monte Carlo (MCMC), using the publicly available code {\sc cosmomc}\footnote{\tt https://cosmologist.info/cosmomc/} \cite{2002PhRvD..66j3511L}. We explore the parameter space of the model signal for a chosen prior range of the parameters $\bm{\mu}$. For the MCMC analysis, we estimate the best fit parameter value $\bm{\hat{\mu}}$ and the confidence intervals. We note that the best-fit parameter $\bm{\hat{\mu}}$ refers to the one that maximizes the log-likelihood. We also use a quantity which gives an essence of the strength of the detected signal, the signal-to-noise ratio, defined in \cite{1992PhRvD..46.5236F} as,
\begin{equation}
    \text{SNR} = \Bigg[\frac{1}{\sigma^2_{\text{rms}}} \int d^2 U \int d \nu \rho_B (\vec{U}, \nu) |S_f(\vec{U},\nu,\bm{\hat{\mu}})|^2 \Bigg]^{1/2}.
    \label{eq:snr}
\end{equation}

In this study, we assume that $\sigma_{\text{rms}}$ does not depend on the $uv$ coverage of the relevant angular or frequency scales. Accounting for the non-uniform $uv$ coverage in the mock observation setup changes $\sigma_{\text{rms}}$ by factors of 2.2, 5.8, and 0.8 for the AA4, AA$\star$, and AA2 configurations of SKA-Low, respectively. Notably, unlike the other two layouts, the AA2 configuration shows a decrease in $\sigma_{\text{rms}}$, which arises from the additional smoothing of the noise map due to the absence of large-scale modes. In practice, incorporating these correction factors would require only a marginal increase in the observation time to achieve the same SNR. We note that, unless stated otherwise, we assume uniform $uv$ coverage throughout this paper.

\subsection{Filter $S_f$}
\label{sec:ft}
In the matched filter formalism, the maximum SNR requires the smoothing filter $S_f$ to be the same as the signal $S(\vec{U},\nu)$. Unfortunately, the spatial distribution of the expected signal is complex and thus, characterizing $S(\vec{U},\nu)$ in terms of a few parameters is not straightforward. The ionized region around a rare bright source, such as a Quasar, is expected to be spherical\footnote{Although quasars emit radiation in all directions, the emission is not strictly isotropic in many physical models. Nevertheless, as a reasonable first approximation, several earlier studies \cite{2012MNRAS.426.3178M, Ross2019} considered isotropic emissions from the quasars.} around the centre of the source. Due to this, earlier studies such as \cite{kanan2007MNRAS.382..809D, 2012MNRAS.426.3178M, mishra2024} considered a spherical top-hat filter. Our previous work \cite{2020MNRAS.496..739G} used a set of five parameters $\left\{R,\theta_X,\theta_Y,\Delta\nu, A_{\TB} \right\}$ where $R$ is radius of the spherical top-hat filter, $A_{\TB}$ is the amplitude of the signal outside the \HII\ bubble. $\theta_X$ and $\theta_Y$  are the x and y-coordinates of the centre of the filter along the angular direction, respectively. $\Delta\nu$ is the frequency difference between the channel that contains the filter's centre and the central frequency channel of the radio observation. 

\begin{figure}
    \centering
  \includegraphics[width=0.75\textwidth]{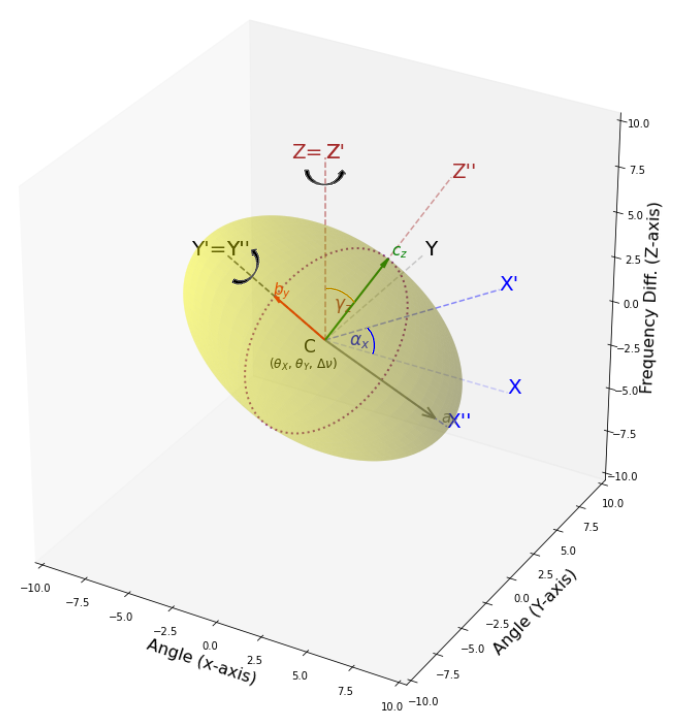}
    \caption{\label{fig:spheroidal_filter_depiction}Visual representation of the spheroid used in this work. The semi-principal axis length $c_Z$ shown in the illustration is for visualization purposes only; in our calculations, it is set equal to $b_Y$. The centre of the spheroid is indicated as $C(\theta_X, \theta_Y, \Delta\nu)$. The original axes are labeled X, Y, and Z. After the first rotation by an angle $\alpha_x$, the axes transform to $X'$, $Y'$, and $Z'$. A subsequent rotation by $\gamma_Z$ results in the final axes, denoted $X''$, $Y''$, and $Z''$, all of which are marked in the figure. A purple dashed great circle is also shown to illustrate the cross-sectional geometry of the spheroid.}
\end{figure}

However, it should be realized that the morphology of the large ionized regions is very unlikely to be spherical in most cases. So, the use of a spherical filter might not be the optimum approach to enhance the detectability of that ionized region from 21 cm observations. Here we introduce a spheroidal filter, which needs more parameters due to the breaking of the radial symmetry of the filter. As two of the three semi-principal axes of the filter have the same length, only two Euler rotations are sufficient to present all rotated configurations of the filter. Figure \ref{fig:spheroidal_filter_depiction} provides a depiction of the spheroidal filter that we use in this work. The parameters used in our filter are listed as follows:

\begin{itemize}
    \item $a_X$ (Mpc): length of the semi-principal axis of the spheroid corresponding to the $x$-axis.
    \item $b_Y$ (Mpc): length of the semi-principal axis of the spheroid corresponding to the $y$-axis.
    \item $\alpha_X$ (radian): rotation angle of the $xy$ plane of the filter around the $z$ axis.
    \item $\gamma_Z$ (radian): rotation angle of the $xz$ plane of the filter around the $y$ axis.
    \item $A_{\TB}$ (mK): amplitude of the signal outside the filter. The signal is set to zero inside the filter.
    \item $\theta_X$ (arcmin): $x$-coordinate of the centre of the spheroidal filter along the angular direction.
    \item $\theta_Y$ (arcmin): $y$-coordinate of the centre of the spheroidal filter along the angular direction.
    \item $\Delta \nu$ (MHz): frequency difference between the central frequency channel of the radio observation and the channel that contains the centre of the filter.
\end{itemize}
We define the set of eight parameters as $\bm{\mu} \equiv \left\{a_X, b_Y, \alpha_X, \gamma_Z, A_{\TB}, \theta_X, \theta_Y, \Delta \nu \right\}$ for convenience. We denote the Fourier transform of the spheroidal filter as $S_f(\vec{U}, \nu; \bm{\mu})$. 

The prior ranges of the filter parameters for the Bayesian analysis could be determined from the observation conditions. Here we choose the ranges as $a_X:[0, 140]$, $b_Y:[0, a_X]$, $\alpha_X:[0, \pi]$, $\gamma_Z:[0,\pi/2]$, $A_{\TB}:[0, 50]$, $\theta_X:[-60, 60]$, $\theta_Y:[-60,60]$ and $\Delta \nu:[-9,9]$.

\section{Results}
\label{sec:res}
First, we study the effect of using the spheroidal filter on a known test ionized region and compare its performance with that of a spherical filter. Thereafter, we apply the spheroidal filter-based framework on the simulated visibilities as described in Section \ref{sec:obs}.

\begin{figure}
    \centering
  \includegraphics[width=\textwidth]{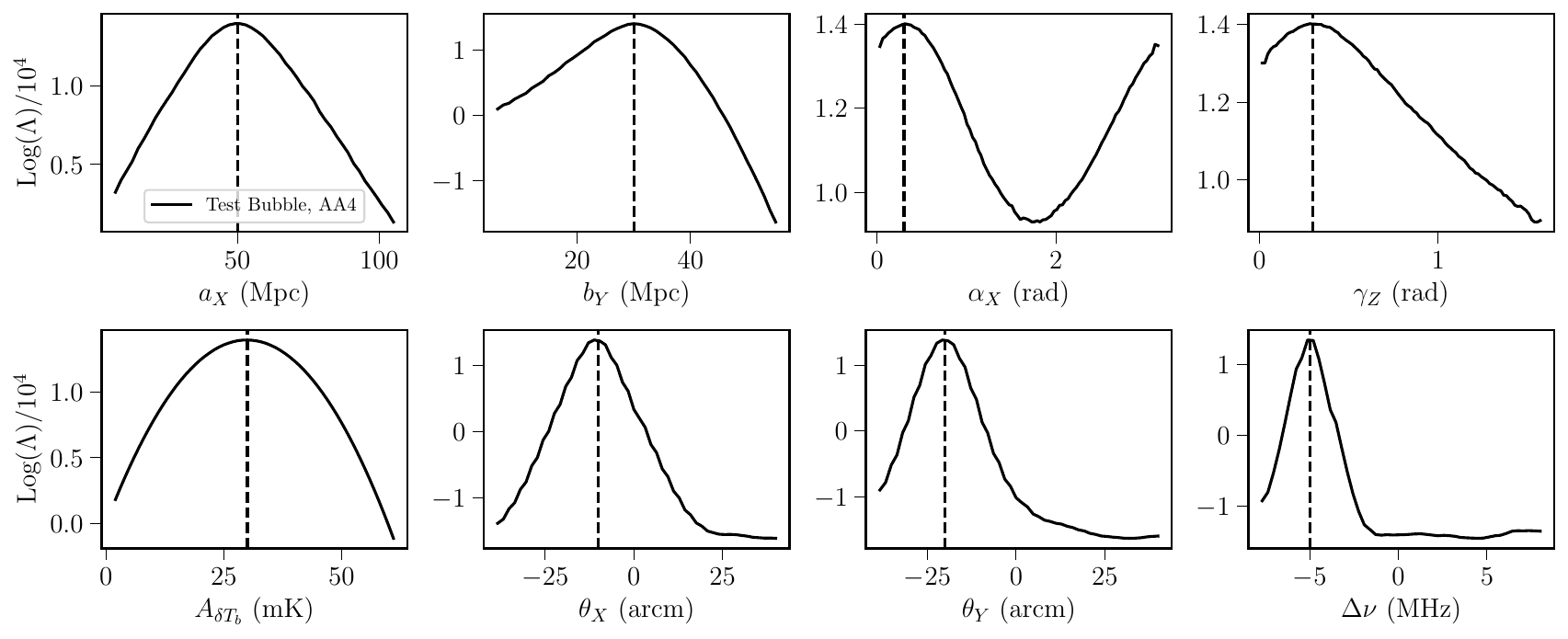}
    \caption{\label{fig:spheroid-config-comparison} Log-likelihood, $\Lambda(\bm{\mu})$, as a function of the spheroidal filter parameters for a test bubble with input values $(a_X, b_Y, \alpha_X, \gamma_Z, A_{\TB}, \theta_X, \theta_Y, \Delta\nu) = (50, 30, 0.3, 0.3, 30, -10, -20, -5)$. In each panel, one parameter is varied while the others are fixed to their input values. The results correspond to a 20 h SKA-low AA4 observation at 177 MHz. Vertical dashed lines denote the true input values in each subplot.}
\end{figure}

\begin{figure}
    \centering
  \includegraphics[width=0.7\textwidth]{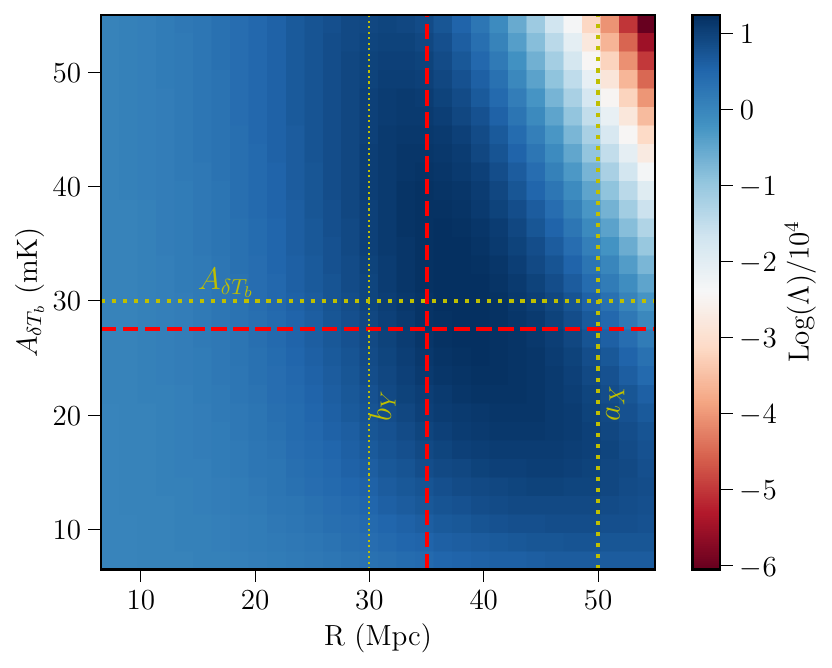}
    \caption{\label{fig:spherical} Log-likelihood, $\Lambda(\bm{\mu})$, as a function of the spherical filter parameters $R$ and $A$. The test case corresponds to a spheroidal ionized bubble with input parameters $(a_X, b_Y, \alpha_X, \gamma_Z, A_{\TB}, \theta_X, \theta_Y, \Delta\nu) = (50, 30, 0.3, 0.3, 30, -10, -20, -5)$. The spherical filter is centered on the bubble, while only $R$ and $A$ are varied. Results are shown for the SKA-low AA4 configuration. The green dotted lines mark the true input values, whereas the red lines indicate the $(R, A)$ values that maximize the likelihood.}
\end{figure}

\begin{table}
\small
    \centering
    \begin{tabular}{|c | c |c|c|c|c|c|c|c|}
    \hline
        Parameters & $\frac{a_X}{\rm Mpc}$  & $\frac{b_Y}{\rm Mpc}$  & $\frac{\alpha_X}{\rm radian}$  & $\frac{\gamma_Z}{\rm radian}$  & $\frac{A_{\TB}}{\rm mK}$ & $\frac{\theta_X}{\rm arcmin}$ & $\frac{\theta_Y}{\rm arcmin}$ & $\frac{\Delta\nu}{\rm MHz}$ \\
        \hline
        \hline
       Inputs (Test case)  & 50 & 30  & 0.3  & 0.3  & 30  & -10  & -20 & -5 \\
       \hline
       Explored range & [0, 140] & [0, 140] & [0, $\pi$] & [0, $\pi/2$] & [0, 50] & [-60, 60] & [-60, 60] & [-9, 9] \\
       \hline
        Best-fit $\bm{\hat{\mu}}$ & 49.98 & 30.03 & 0.31 & 0.28  & 29.62  & -9.99 & -20.02 & -4.98  \\
        \hline
    \end{tabular}
    \caption{Bayesian analysis results for the test spheroidal bubble located at redshift 7. This analysis is done using $20$ h of the SKA-low AA4 mock data. }
    \label{tab:testcase}
\end{table}

\begin{figure}[]
\centering 
\includegraphics[width=\textwidth]{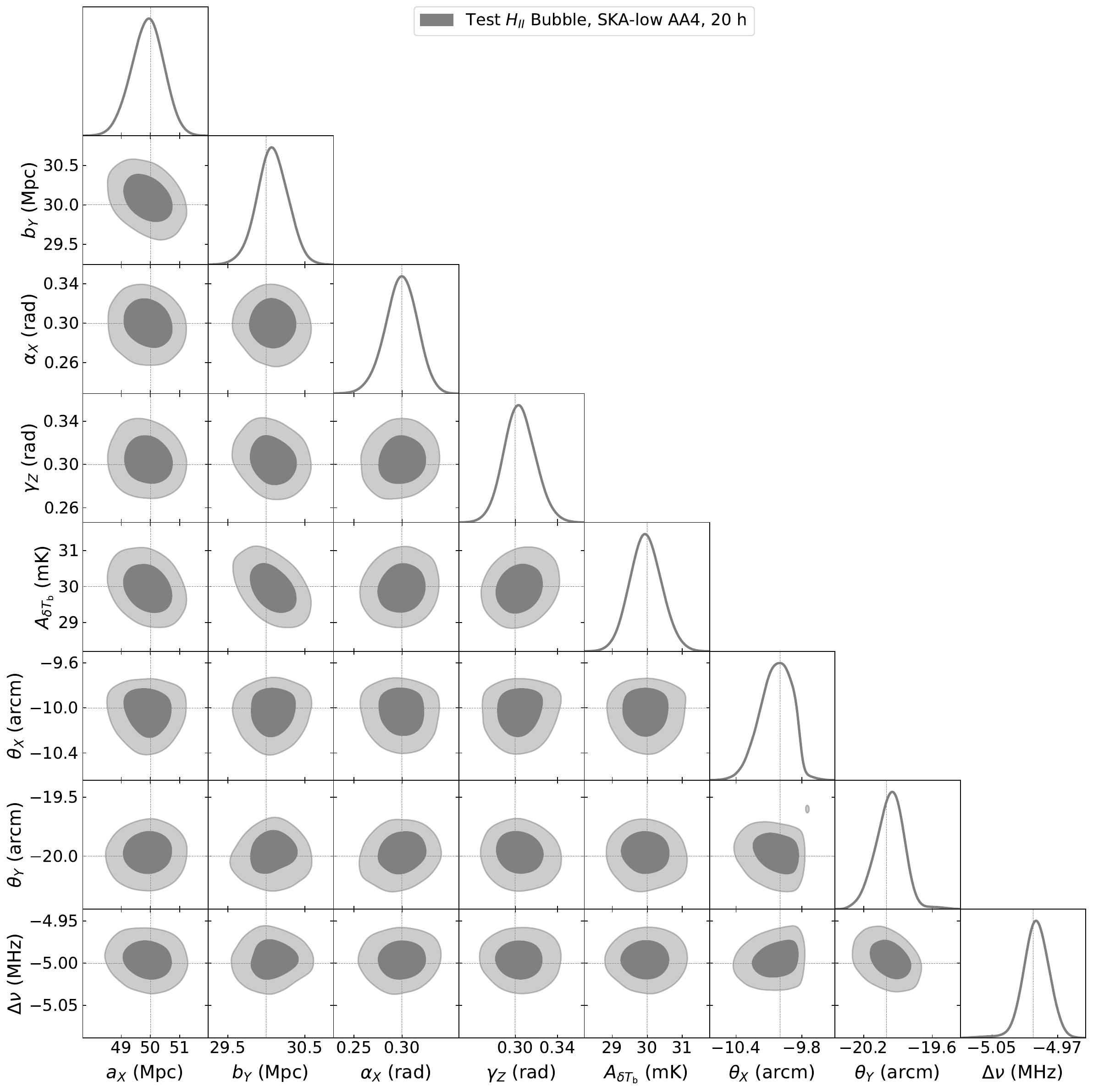}
\caption{\label{fig:test-cases-spheroid-corner-results} Posterior distributions of the spheroidal bubble parameters from mock observations of the test bubble. The two-dimensional contours denote the $1\sigma$ and $2\sigma$ confidence levels, while the diagonal panels show the marginalized one-dimensional distributions. The analysis assumes $20$ h of SKA-low AA4 observations at $z=7$. Vertical dotted lines mark the input signal parameters.}
\end{figure}

\subsection{Test Case: isolated spheroidal ionized region in uniform background}
\label{sec:testcase}
We present a test case to demonstrate the performance of our spheroidal filter in a simplified ionization scenario. Specifically, we simulate an isolated spheroidal ionized region with parameters $a_X=50$ Mpc, $b_Y=30$ Mpc, $\alpha_X = 0.3$ rad, $\gamma_Z=0.3$ rad, $A_{\TB}=30$ mK, $\theta_X=-10$ arcmin, $\theta_Y=-20$ arcmin, and $\Delta \nu=-5$ MHz. The center of the \HII\ region is assumed to lie at $z=7$. For this demonstration, we adopt a $20$ h observation using the SKA-low AA4 configuration.

We begin by varying one filter parameter at a time while keeping the remaining parameters fixed to the values of the test input spheroidal bubble. Figure \ref{fig:spheroid-config-comparison} illustrates how the log-likelihood changes as a function of these parameters. As expected, the likelihood peaks when the filter parameters coincide with those of the input spheroidal bubble. In this case, the maximum log-likelihood and the corresponding SNR are found to be $1.4\times 10^4$ and $117$, respectively.

Next, we apply our previously developed spherical filter framework to the simulated spheroidal bubble signal. In this case, we fix the filter parameters $\theta_X$, $\theta_Y$, and $\Delta \nu$ to their known input values, while varying only $R$ and $A_{\TB}$. Here, $R$ denotes the radius of the spherical filter (in Mpc), and $A_{\TB}$ represents the signal amplitude outside the filter. Figure \ref{fig:spherical} shows the resulting log-likelihood distribution as a function of $R$ and $A_{\TB}$. The maximum log-likelihood and SNR obtained are $1.2\times 10^4$ and $108$, respectively, both lower than the corresponding values achieved with the spheroidal filter. The best-fit parameters are $R=35$ Mpc and $A_{\TB}=27.5$ mK. The inferred radius lies between the input values of $b_Y$ and $a_X$, while the recovered $A_{\TB}$ is smaller than the true input, reflecting compensation for the non-zero signal within $R$.

Next, we explore the parameter space of the spheroidal filter for the test case using the SKA-low AA4 configuration. Figure \ref{fig:test-cases-spheroid-corner-results} presents the posterior distributions of the filter parameters obtained from the Bayesian analysis. The one-dimensional PDFs demonstrate that the input parameter values are recovered with high accuracy. The best-fit values, $\bm{\hat{\mu}}$, are summarized in Table \ref{tab:testcase}, and closely match the real inputs. The corresponding SNR calculated from these best-fit values of the parameters is $117$.


\subsection{Realistic Scenarios}
\label{sec:real_scen}

\begin{figure}[]
\centering 
\includegraphics[width=\textwidth]{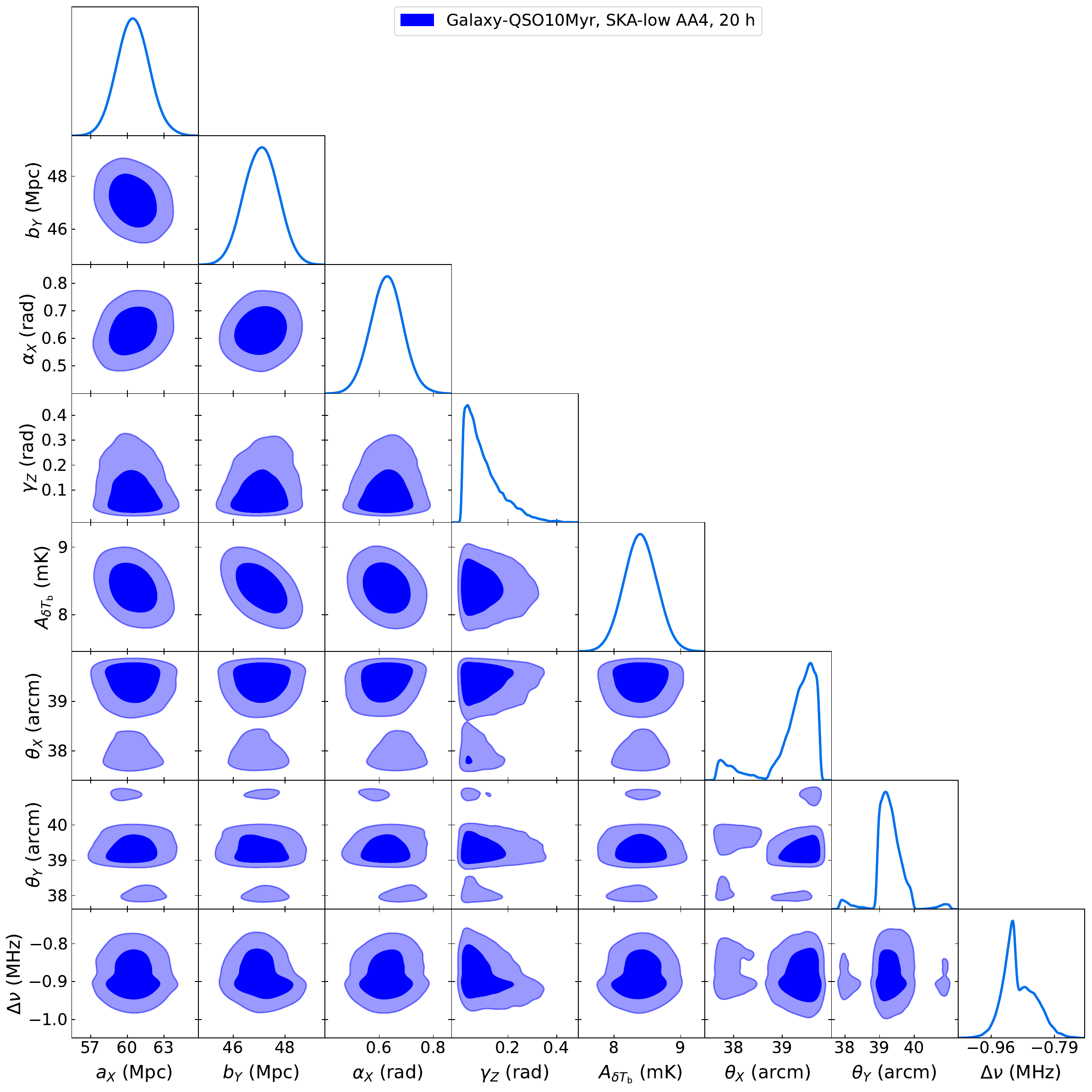}\llap{\raisebox{8.5cm}{\includegraphics[height=6.4cm, width=7.27cm]{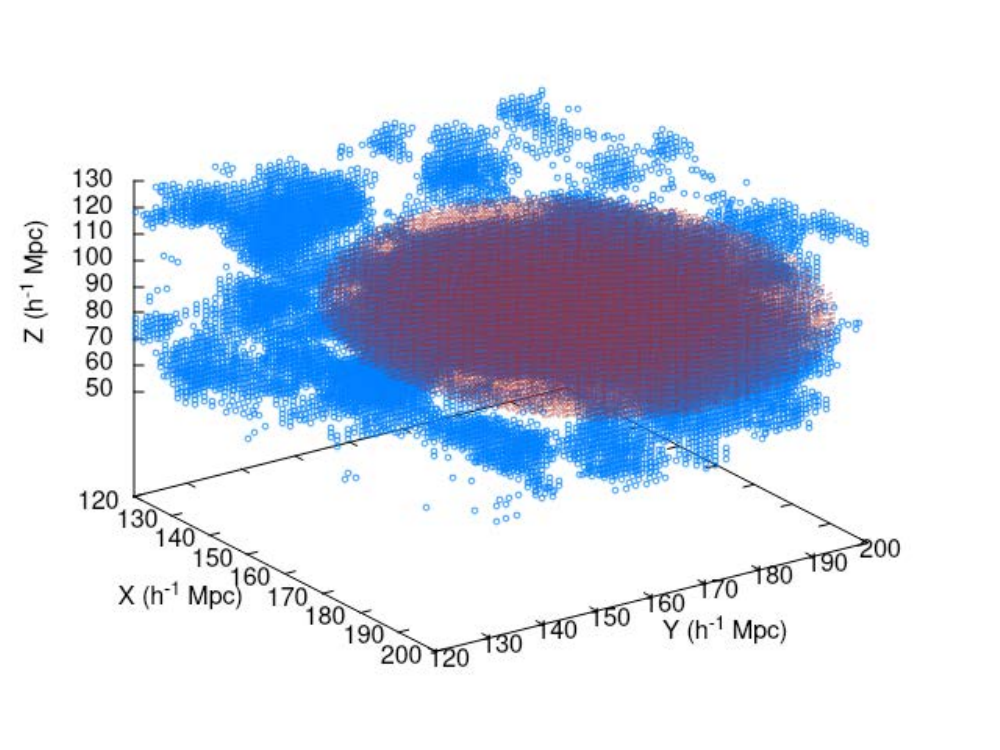}}}
\caption{\label{fig:sph_5D10param8_tri} Posterior distribution of the spheroidal filter parameters for the \textit{Galaxy-QSO10Myr} 21 cm signal scenario. The analysis is based on a 20 h mock observation with the SKA-low AA4 configuration. The two-dimensional contours correspond to the $1\sigma$ and $2\sigma$ confidence levels, while the diagonal panels show the marginalized one-dimensional probability distributions of each parameter. The inset illustrates the three-dimensional ionized region (blue) from the simulation and the best-fit spheroidal filter (red) recovered in this scenario.}
\end{figure}

\begin{figure}[]
\centering 
\includegraphics[width=\textwidth]{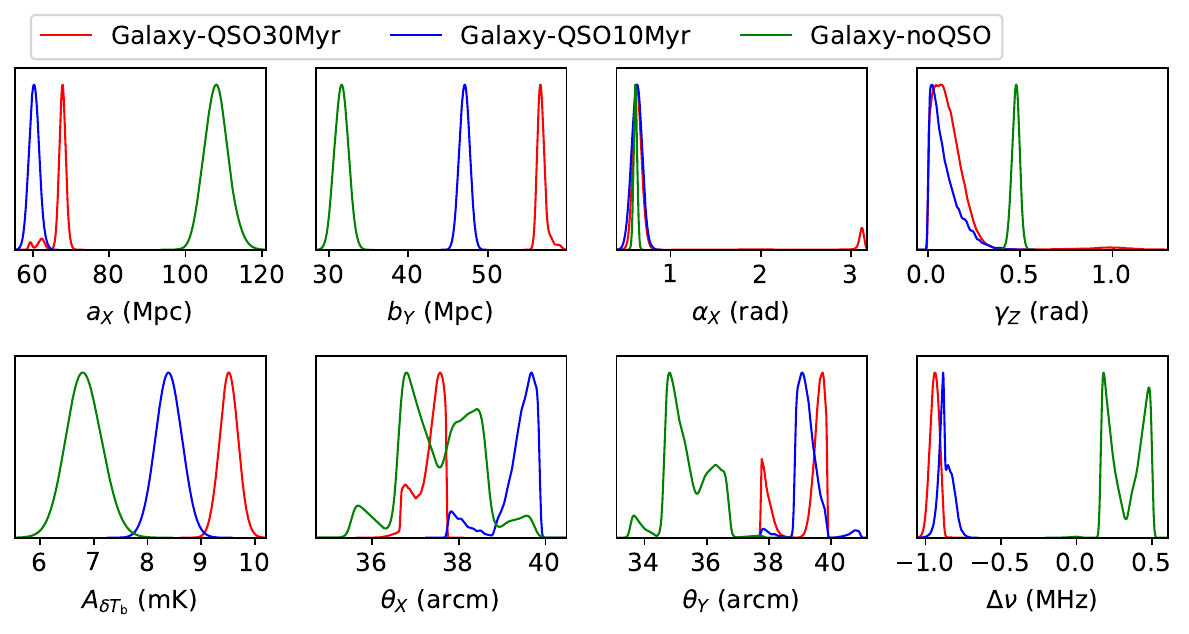}
\caption{\label{fig:QSOparam8_1D} Marginalized probability distributions of the spheroidal filter parameters for the three realistic 21 cm signal scenarios considered in this study. The red, blue, and green curves correspond to the \textit{Galaxy-QSO30Myr}, \textit{Galaxy-QSO10Myr}, and \textit{Galaxy-noQSO} cases, respectively. The results are derived from a 20 h mock observation at $z=7$ using the SKA-low AA4 configuration.  }
\end{figure}

\begin{table}
\small
    \centering
    \renewcommand\arraystretch{1.2}
    \begin{tabular}{|c | c |c|c|c|c|c|c|c|}
    \hline
        Parameters & $\frac{a_X}{\rm Mpc}$  & $\frac{b_Y}{\rm Mpc}$  & $\frac{\alpha_X}{\rm radian}$  & $\frac{\gamma_Z}{\rm radian}$  & $\frac{A_{\TB}}{\rm mK}$ & $\frac{\theta_X}{\rm arcmin}$ & $\frac{\theta_Y}{\rm arcmin}$ & $\frac{\Delta\nu}{\rm MHz}$ \\
        \hline
        \hline
         Explored range & [0, 140] & [0, 140] & [0, $\pi$] & [0, $\pi/2$] & [0, 50] & [-60, 60] & [-60, 60] & [-9, 9] \\
         \hline\hline
        Best-fit (10Myr-AA4) & 60.29 & 47  & 0.62  & 0.006  & 8.34  & 39.65  & 39.14  & -0.84 \\
        \hline
        Best-fit (30Myr-AA4) & 67.5 & 56.4 & 0.62  & 0.08  & 9.58  & 37.61  & 39.65  & -0.92 \\
        \hline
        Best-fit (noQSO-AA4) & 110.21 & 31.66  & 0.61  & 0.45  & 6.94 & 36.74 & 35.15 & 0.49 \\
        \hline
        Best-fit (10Myr-AA$^{\star}$) & 60.92 & 46.12  & 0.63  & 0.13  & 8.67  & 39.77  & 40.8  & -0.92 \\
        \hline
        Best-fit (10Myr-AA2) & 134 & 45.3 & 0.000 & 0.63  & 6.71  & 52.1 & 32.62 & -2.3    
        \\
        \hline 
    \end{tabular}
    \caption{Bayesian analysis results for the $Galaxy$–$QSO10$Myr, $Galaxy$–$QSO30$Myr, and $Galaxy$–$noQSO$ scenarios. The MCMC analysis is based on SKA-low configurations with 20, 55 and 1100 h mock observations for the AA4, AA$\star$ and AA2 SKA-low configurations, respectively. The quasar position in the simulation is fixed at $(\theta_X = 34.94', \theta_Y = 37.625', \Delta\nu = -0.875$ MHz).}
    \label{tab:qsocaseAA4}
\end{table}

\begin{table}[h!]
    \centering
    \caption{Signal-to-noise ratio achieved from EoR scenarios considered in this study. From left to right, different columns represent the considered EoR scenarios, SKA layouts, total observation hours, and the SNR for considering uniform and realistic non-uniform $uv$ coverage in $\sigma_{\rm rms}$ estimates, respectively.}
    \label{tab:snr}
    \begin{tabular}{|c|c|c|c|c|}
        \hline
        EoR Scenario & SKA layout & $t_{\rm obs}$ (h) & SNR (uniform $uv$) & SNR (non-uniform $uv$) \\
        \hline
        Galaxy-QSO10Myr & AA4 & 20 & 60 & 27 \\
        \hline
        Galaxy-QSO30Myr & AA4 & 20 & 89 & 40 \\
        \hline
        Galaxy-noQSO & AA4 & 20 & 44 & 20 \\
        \hline
        Galaxy-QSO10Myr & AA$\star$ & 55 & 76 & 13 \\
        \hline
        Galaxy-QSO10Myr & AA2 & 1100 & 210 & 255 \\
         \hline
    \end{tabular}
\end{table}

We begin with the \textit{Galaxy-QSO10Myr} model (see Section~\ref{sec:simdtb}), which includes an ionized region produced by a quasar with a lifetime of $10$ Myr. The corresponding posterior distributions are shown in Figure~\ref{fig:sph_5D10param8_tri}. In this analysis, we performed a blind search, treating the bubble centre as a free parameter in the MCMC sampling. This allows the filter to adjust to the optimal configuration, unlike the case where the centre is fixed to the quasar’s location. The best-fit parameters are $\bm{\hat{\mu}}=\left\{60.29, 47, 0.62, 0.006, 8.34, 39.65, 39.14, -0.84\right\}$ (see Table~\ref{tab:qsocaseAA4}). Notably, the inferred bubble centre does not coincide with the quasar position in the simulation ($\theta_X = 34.94'$, $\theta_Y = 37.625'$, $\Delta \nu = -0.875$ MHz), suggesting that the ionized region grows asymmetrically around the quasar. The best-fit value of $a_X$ is significantly larger than $b_Y$, confirming the non-spherical morphology of the region. Furthermore, the orientation of the recovered spheroid matches well with the ionized structure surrounding the most massive dark matter halo (see inset of Figure~\ref{fig:sph_5D10param8_tri}). The SNR from this analysis is 60, an improvement over the SNR of 50 reported by \cite{2020MNRAS.496..739G} when applying a spherical filter to the same mock dataset.

Next, we analyze the remaining two scenarios, \textit{Galaxy-noQSO} and \textit{Galaxy-QSO30Myr}, as described in Section~\ref{sec:simdtb}. The marginalized probability distributions of the filter parameters for all three realistic cases are shown in Figure~\ref{fig:QSOparam8_1D}, while the corresponding best-fit values are summarized in Table~\ref{tab:qsocaseAA4}. For the \textit{Galaxy-noQSO} scenario, the best-fit value of $a_X$ is the largest, whereas $b_Y$ is the smallest, suggesting that the dominant ionized region has an elongated or tunnel-like structure in the absence of a bright quasar. In contrast, the \textit{Galaxy-QSO30Myr} case yields the largest best-fit value of $b_Y$, implying the widest ionized region among the three scenarios. The multiple peaks in the marginalized distributions of the position parameters further indicate that the morphology of the largest ionized region is more complex than a simple spheroidal shape. The SNR values obtained for the \textit{Galaxy-noQSO} and \textit{Galaxy-QSO30Myr} scenarios are $44$ and $89$, respectively. For comparison, \cite{2020MNRAS.496..739G} reported SNR values of $32$ and $76$ for the same two scenarios when applying a spherical filter.

For all three reionization scenarios, the marginalized PDFs of $\alpha_X$ and $\gamma_Z$ reveal a preferred orientation of the best-fit spheroidal filter. This is consistent with the expectation that the largest ionized regions exhibit a tunnel-like morphology. Consequently, the best-fit values of $\alpha_X$ and $\gamma_Z$ provide valuable information about the orientation of these extended ionized structures.

\subsubsection{Implications for SKA-low configurations}
\label{sec:ska_im}

\begin{figure}
\centering 
\includegraphics[width=\textwidth]{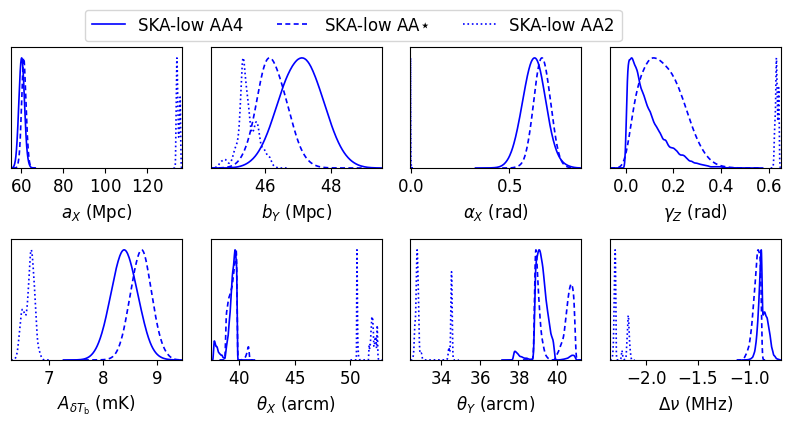}
\caption{\label{fig:sph_5_all_cases_tri_SKA} Marginalized probability distributions of the filter parameters for the fiducial 21 cm signal scenario, \textit{Galaxy-QSO10Myr}. The solid, dashed, and dotted curves correspond to the SKA-low AA4, AA$^{\star}$, and AA2 configurations, respectively. For the MCMC analysis, we assume $20$ h of observation time.}
\end{figure}

Next, we investigate how our findings change for considering the other two SKA-low configurations. We perform MCMC analysis on the fiducial EoR scenario while considering the SKA-low-AA$^{\star}$ and SKA-low-AA2 configurations. We note that the value of $t_{\rm obs}$ is $20$ h, $55$ h and $1100$ h for the AA4, AA$^{\star}$, and AA2 SKA-low configurations, respectively. These are chosen considering that SNR$ \propto \sqrt{t_{\rm obs}N^2_{\rm ant}}$. This, however, does not mean the achieved SNR for these different cases will be the same as the SNR also depends on $\rho_B (\vec{U}, \nu)$ (Equ. \ref{eq:snr}).

Figure \ref{fig:sph_5_all_cases_tri_SKA} shows the marginalized PDFs of the filter parameters as obtained from the MCMC analysis for considering three different SKA-low configurations. The best-fit values of the filter parameters are presented in Table~\ref{tab:qsocaseAA4}. These are close to each other for the AA4 and AA$^\star$ configurations, indicating that the location and the properties of the largest ionized region can be better detected using those SKA-low configurations. The best-fit parameters from the analysis that used the AA2 configuration do not indicate the same region preferred by the other analysis. The reason might be that the largest ionized region is smoothed out and formed a very different structure which is shown in Figure \ref{fig:signal}.

The corresponding SNRs of this EoR scenario while using the AA4, AA$^\star$, and AA2 layouts are approximately 60, 76, and 210, respectively. The higher SNR values in the AA$^\star$ and AA2 configurations arise because their corresponding $\rho_B(\vec{U}, \nu)$ distributions span a narrower range of baseline modes. In our framework, the SNR scales as $\sqrt{t_{\rm obs}N_{\rm ant}^2}$. Consequently, a $10\sigma$ detection of the largest ionized region in the fiducial reionization model would require roughly $\sim 0.5$, $1$, and $3$ hours of observation for the AA4, AA$^\star$, and AA2 configurations of SKA-Low, respectively. 

In table \ref{tab:snr}, we have listed the SNRs achieved from different scenarios considered in this study. We note that these estimates are rather optimistic, as they do not account for $\sigma_{\rm rms}$ corrections arising from non-uniform baseline distributions, data-loss effects, or residual imaging artefacts during the analysis pipeline. Table \ref{tab:snr} also presents revised SNRs after considering non-uniform $uv$ coverage through the estimation of $\sigma_{\rm rms}$. This step changes the SNRs by factors of 2.2, 5.8, and 0.8 for the AA4, AA$^\star$, and AA2 configurations of SKA-Low, respectively.

\section{Conclusions \& Discussion}
\label{sec:con}
In this work, we have employed a matched-filter-based technique for optimally detecting and constraining the properties of large ionized regions during the cosmic reionization. We used different mock observations with upcoming configurations of the SKA-low radio interferometer for simulated 21 cm signal light-cones. Here, we have considered AA2, AA$^{\star}$ and AA4 configurations of the SKA-low with $68$, $307$ and $512$ antennas, respectively.

Here, we have considered the previously introduced Bayesian approach \cite{2020MNRAS.496..739G} which is well-suited for detecting large ionized regions in both targeted and blind searches in the Epoch of Reionization, depending on whether the location of the bubble is known. In this study, we have used a spheroidal filter in the Bayesian framework instead of a spherical filter as used in \cite{2020MNRAS.496..739G}. The spheroidal top-hat filter is characterized by eight parameters: the three position coordinates of the centre of the spheroid ($\theta_X, \theta_Y, \Delta\nu$), an amplitude that measures the contrast between the signal outside and inside the bubble ($A_{\TB}$), two semi-principal axes of the spheroid ($a_X, b_Y$), and two parameters that dictate the angular orientation of the spheroid ($\alpha_X, \gamma_Z$). 

We first validated our framework using a controlled test case, where a spheroidal ionized region with predefined parameters was simulated and analyzed. The framework successfully recovered the position, size, orientation, and average brightness temperature of the region. For a spheroidal bubble with $a_X=50$ Mpc and $b_Y=30$ Mpc at $z=7$, the Bayesian analysis with a spheroidal filter on $20$ h SKA-low AA4 mock observations yielded an SNR of $117$. By contrast, applying a spherical filter to the same signal resulted in a slightly lower SNR of $108$.

We next applied the framework to the same realistic ionization maps as used in \cite{2020MNRAS.496..739G}, and found that the results do explain the characteristics of the largest ionized region more accurately than employing a spherical filter.

We find that, when applying the spheroidal filter to $20$ h of mock observations with SKA-low AA4 for the galaxy-driven reionization scenario, the SNR of the largest ionized region at redshift $7$ is $44$. This is $\sim 11$ higher than the SNR obtained using a spherical filter on the same mock data.

The SNR of the largest ionized region surrounding a bright quasar with properties consistent with \cite{Mortlock11} is found to be 89 for $20$ h of observation with SKA-low AA4. The SNR depends on the size of the ionized region, which in turn is determined by the properties of the quasar. When employing a spherical filter within the Bayesian framework, the SNR decreases to $76$. Across all reionization scenarios explored in this study, we consistently find that using a spheroidal filter yields a higher SNR compared to the spherical filter.

The marginalized PDFs of $\alpha_X$ and $\gamma_Z$ exhibit a preferred orientation of the best-fit spheroidal filter across all reionization scenarios considered. These best-fit values provide information about the orientation of the largest ionized regions.

We find that the information on ionized regions is limited with the SKA-low AA2 configuration. Nevertheless, a $10\sigma$ detection of the largest ionized region in a typical reionization scenario can be achieved with $\sim 0.5$, $1$, and $3$ h of observation using the AA4, AA$^{\star}$, and AA2 configurations, respectively.

For a quasar-driven region, the parameters recovered from our analysis carry direct physical significance. The inferred position coordinates identify the location of the largest ionized region that hosts the quasar. The semi-principal axis lengths and associated angular parameters provide valuable information about the surrounding galaxy distribution, helping to disentangle the relative contributions of clustered galaxies and the luminous quasar to the observed ionization structure.

In the absence of a bright quasar, the morphology of the largest ionized region is expected to be more irregular. This reduces the contrast between the signal inside the filter volume and the surrounding medium. The maximum-likelihood solution is obtained when this contrast matches the parameter $A_{\TB}$. Therefore, the small value of $A_{\TB}$ obtained for the \textit{Galaxy-noQSO} scenario suggests that the ionized structure deviates significantly from a simple spheroidal form. Nevertheless, the presence of large individual ionized regions in this case points to strong clustering among early galaxies.

Overall, the parameters derived through our formalism highlight regions that are promising targets for follow-up observations at infrared wavelengths, enabling further investigation of the properties of these early ionizing sources. Applying blind search techniques in this context could yield valuable insights into the locations of the first luminous sources at very high redshifts, thereby providing important guidance for near-infrared observations with the James Webb Space Telescope (JWST). Although the present work primarily focuses on detecting ionized regions during the EoR, the same methodology can be extended to identify emission structures from the Cosmic Dawn \cite{ghara15a, ghara16, Ross2019}. 

Our analysis demonstrates that the detectability of large ionized regions during the EoR can be substantially improved by employing a spheroidal filter rather than a spherical filter on the observed visibilities. Nevertheless, these results rely on several simplifying assumptions. We have assumed an idealized case in which visibilities are perfectly calibrated and free of residual artefacts - a condition that is difficult to achieve in practice. Moreover, our analysis presumes that both Galactic and extragalactic foregrounds are spectrally smooth and can be subtracted with high accuracy. In reality, handling the complex visibilities will inevitably introduce additional uncertainties beyond those considered here. Thus, the predictions presented in this work should be regarded as optimistic.

The matched filtering technique and the traditional imaging technique contribute in distinct ways to the detection of the EoR 21-cm signal. The first one is ideally suited for extracting weak signals when their shapes are reasonably well known, allowing it to reach high SNR with comparatively little observation time. In contrast, traditional imaging methods, although less sensitive in low-SNR conditions, provide a template-free reconstruction of the sky, making it valuable for examining the irregular and complex structure of ionized regions. When combined, these approaches complement each other. The matched filtering technique serves as an efficient detection tool, while imaging offers the broader physical insight necessary for interpreting the detected features.

\acknowledgments
RG acknowledges support from SERB, DST Ramanujan Fellowship no. RJF/2022/000141. AM  acknowledges financial support from Council of Scientific and Industrial Research (CSIR) via  CSIR-SRF fellowships under grant no. 09/0096(13611)/2022-EMR-I.


\bibliographystyle{JHEP}
\bibliography{mybib}

@ARTICLE{2020Mevius,
       author = {{Mevius}, M. and {Mertens}, F. and {Koopmans}, L.~V.~E. and {Offringa}, A.~R. and {Yatawatta}, S. and {Brentjens}, M.~A. and {Chapman}, E. and {Ciardi}, B. and {Gan}, H. and {Gehlot}, B.~K. and {Ghara}, R. and {Ghosh}, A. and {Giri}, S.~K. and {Iliev}, I.~T. and {Mellema}, G. and {Pandey}, V.~N. and {Zaroubi}, S.},
        title = "{A numerical study of 21-cm signal suppression and noise increase in direction-dependent calibration of LOFAR data}",
      journal = {\mnras},
         year = 2022,
        month = jan,
       volume = {509},
       number = {3},
        pages = {3693-3702},
          doi = {10.1093/mnras/stab3233},
archivePrefix = {arXiv},
       eprint = {2111.02537},
 primaryClass = {astro-ph.CO},
       adsurl = {https://ui.adsabs.harvard.edu/abs/2022MNRAS.509.3693M}
}

@ARTICLE{2020MNRAS.493.4711T,
       author = {{Trott}, Cathryn M. and {Jordan}, C.~H. and {Midgley}, S. and
         {Barry}, N. and {Greig}, B. and {Pindor}, B. and {Cook}, J.~H. and
         {Sleap}, G. and {Tingay}, S.~J. and {Ung}, D. and {Hancock}, P. and
         {Williams}, A. and {Bowman}, J. and {Byrne}, R. and {Chokshi}, A. and
         {Hazelton}, B.~J. and {Hasegawa}, K. and {Jacobs}, D. and
         {Joseph}, R.~C. and {Li}, W. and {Line}, J.~L.~B. and {Lynch}, C. and
         {McKinley}, B. and {Mitchell}, D.~A. and {Morales}, M.~F. and
         {Ouchi}, M. and {Pober}, J.~C. and {Rahimi}, M. and {Takahashi}, K. and
         {Wayth}, R.~B. and {Webster}, R.~L. and {Wilensky}, M. and
         {Wyithe}, J.~S.~B. and {Yoshiura}, S. and {Zhang}, Z. and {Zheng}, Q.},
        title = "{Deep multiredshift limits on Epoch of Reionization 21 cm power spectra from four seasons of Murchison Widefield Array observations}",
      journal = {\mnras},
         year = 2020,
        month = apr,
       volume = {493},
       number = {4},
        pages = {4711-4727},
          doi = {10.1093/mnras/staa414},
archivePrefix = {arXiv},
       eprint = {2002.02575},
 primaryClass = {astro-ph.CO},
       adsurl = {https://ui.adsabs.harvard.edu/abs/2020MNRAS.493.4711T}
}

@ARTICLE{2018MNRAS.479.5596G,
       author = {{Giri}, Sambit K. and {Mellema}, Garrelt and {Ghara}, Raghunath},
        title = "{Optimal identification of H II regions during reionization in 21-cm observations}",
      journal = {\mnras},
         year = 2018,
        month = oct,
       volume = {479},
       number = {4},
        pages = {5596-5611},
          doi = {10.1093/mnras/sty1786},
archivePrefix = {arXiv},
       eprint = {1801.06550},
 primaryClass = {astro-ph.CO},
       adsurl = {https://ui.adsabs.harvard.edu/abs/2018MNRAS.479.5596G}
}

@ARTICLE{paciga13,
   author = {{Paciga}, G. and {Albert}, J.~G. and {Bandura}, K. and {Chang}, T.-C. and 
	{Gupta}, Y. and {Hirata}, C. and {Odegova}, J. and {Pen}, U.-L. and 
	{Peterson}, J.~B. and {Roy}, J. and {Shaw}, J.~R. and {Sigurdson}, K. and 
	{Voytek}, T.},
    title = "{A simulation-calibrated limit on the H I power spectrum from the GMRT Epoch of Reionization experiment}",
  journal = {\mnras},
archivePrefix = "arXiv",
   eprint = {1301.5906},
 primaryClass = "astro-ph.CO",
     year = 2013,
    month = jul,
   volume = 433,
    pages = {639-647},
      doi = {10.1093/mnras/stt753},
   adsurl = {http://adsabs.harvard.edu/abs/2013MNRAS.433..639P}
}

@ARTICLE{2024MNRAS.527.3517M,
       author = {{Mertens}, Florent G. and {Bobin}, J{\'e}r{\^o}me and {Carucci}, Isabella P.},
        title = "{Retrieving the 21-cm signal from the Epoch of Reionization with learnt Gaussian process kernels}",
      journal = {\mnras},
         year = 2024,
        month = jan,
       volume = {527},
       number = {2},
        pages = {3517-3531},
          doi = {10.1093/mnras/stad3430},
archivePrefix = {arXiv},
       eprint = {2307.13545},
 primaryClass = {astro-ph.CO},
       adsurl = {https://ui.adsabs.harvard.edu/abs/2024MNRAS.527.3517M}
}

@misc{yatawatta20112011,
  title={2011 XXXth URSI General Assembly and Scientific Symposium},
  author={Yatawatta, S and Banerjee, P},
  year={2011},
  publisher={IEEE Piscataway}
}

@inproceedings{Koopmans_2015, series={AASKA14},
   title={The Cosmic Dawn and Epoch of Reionisation with SKA},
   url={http://dx.doi.org/10.22323/1.215.0001},
   DOI={10.22323/1.215.0001},
   booktitle={Proceedings of Advancing Astrophysics with the Square Kilometre Array — PoS(AASKA14)},
   publisher={Sissa Medialab},
   author={Koopmans, Leon and Pritchard, J and Mellema, G and Aguirre, J and Ahn, K and Barkana, R and van Bemmel, I and Bernardi, G and Bonaldi, A and Briggs, F and de Bruyn, A. G. and Chang, T. C. and Chapman, E and Chen, X and Courty, B and Dayal, P. and Ferrara, A. and Fialkov, A. and Fiore, F and Ichiki, K. and Illiev, I. T. and Inoue, S and Jelic, V and Jones, M and Lazio, J and Maio, U and Majumdar, S and Mack, K. J. and Mesinger, A. and Morales, M F. and Parsons, A. and Pen, U.L. and Santos, M and Schneider, R and Semelin, B and de Souza, R S and Subrahmanyan, R and Takeuchi, T and Vedantham, H and Wagg, J and Webster, R and Wyithe, S and Datta, Kanan Kumar and Trott, C.},
   year={2015},
   month=may, collection={AASKA14} }

@ARTICLE{2024MNRAS.527.7835A,
       author = {{Acharya}, Anshuman and {Mertens}, Florent and {Ciardi}, Benedetta and {Ghara}, Raghunath and {Koopmans}, L{\'e}on V.~E. and {Giri}, Sambit K. and {Hothi}, Ian and {Ma}, Qing-Bo and {Mellema}, Garrelt and {Munshi}, Satyapan},
        title = "{21-cm signal from the Epoch of Reionization: a machine learning upgrade to foreground removal with Gaussian process regression}",
      journal = {\mnras},
         year = 2024,
        month = jan,
       volume = {527},
       number = {3},
        pages = {7835-7846},
          doi = {10.1093/mnras/stad3701},
archivePrefix = {arXiv},
       eprint = {2311.16633},
 primaryClass = {astro-ph.CO},
       adsurl = {https://ui.adsabs.harvard.edu/abs/2024MNRAS.527.7835A}
}

@ARTICLE{kanan2007MNRAS.382..809D,
   author = {{Datta}, K.~K. and {Bharadwaj}, S. and {Choudhury}, T.~R.},
    title = "{Detecting ionized bubbles in redshifted 21-cm maps}",
  journal = {\mnras},
   eprint = {astro-ph/0703677},
     year = 2007,
    month = dec,
   volume = 382,
    pages = {809-818},
      doi = {10.1111/j.1365-2966.2007.12421.x},
   adsurl = {http://adsabs.harvard.edu/abs/2007MNRAS.382..809D}
}

@ARTICLE{2002PhRvD..66j3511L,
       author = {{Lewis}, Antony and {Bridle}, Sarah},
        title = "{Cosmological parameters from CMB and other data: A Monte Carlo approach}",
      journal = {\prd},
         year = "2002",
        month = "Nov",
       volume = {66},
          eid = {103511},
        pages = {103511},
          doi = {10.1103/PhysRevD.66.103511},
archivePrefix = {arXiv},
       eprint = {astro-ph/0205436},
 primaryClass = {astro-ph},
       adsurl = {https://ui.adsabs.harvard.edu/\#abs/2002PhRvD..66j3511L}
}

@ARTICLE{2018Natur.553..473B,
       author = {{Ba{\~n}ados}, Eduardo and {Venemans}, Bram P. and
         {Mazzucchelli}, Chiara and {Farina}, Emanuele P. and {Walter}, Fabian and
         {Wang}, Feige and {Decarli}, Roberto and {Stern}, Daniel and
         {Fan}, Xiaohui and {Davies}, Frederick B. and {Hennawi}, Joseph F. and
         {Simcoe}, Robert A. and {Turner}, Monica L. and {Rix}, Hans-Walter and
         {Yang}, Jinyi and {Kelson}, Daniel D. and {Rudie}, Gwen C. and
         {Winters}, Jan Martin},
        title = "{An 800-million-solar-mass black hole in a significantly neutral Universe at a redshift of 7.5}",
      journal = {\nat},
         year = "2018",
        month = "Jan",
       volume = {553},
        pages = {473-476},
          doi = {10.1038/nature25180},
archivePrefix = {arXiv},
       eprint = {1712.01860},
 primaryClass = {astro-ph.GA},
       adsurl = {https://ui.adsabs.harvard.edu/\#abs/2018Natur.553..473B}
}

@ARTICLE{datta2012a,
   author = {{Datta}, K.~K. and {Friedrich}, M.~M. and {Mellema}, G. and 
	{Iliev}, I.~T. and {Shapiro}, P.~R.},
    title = "{Prospects of observing a quasar H II region during the epoch of reionization with the redshifted 21-cm signal}",
  journal = {\mnras},
archivePrefix = "arXiv",
   eprint = {1203.0517},
 primaryClass = "astro-ph.CO",
     year = 2012,
    month = jul,
   volume = 424,
    pages = {762-778},
      doi = {10.1111/j.1365-2966.2012.21268.x},
   adsurl = {http://adsabs.harvard.edu/abs/2012MNRAS.424..762D}
}

@ARTICLE{mao12,
   author = {{Mao}, Y. and {Shapiro}, P.~R. and {Mellema}, G. and {Iliev}, I.~T. and 
	{Koda}, J. and {Ahn}, K.},
    title = "{Redshift-space distortion of the 21-cm background from the epoch of reionization - I. Methodology re-examined}",
  journal = {\mnras},
archivePrefix = "arXiv",
   eprint = {1104.2094},
 primaryClass = "astro-ph.CO",
     year = 2012,
    month = may,
   volume = 422,
    pages = {926-954},
      doi = {10.1111/j.1365-2966.2012.20471.x},
   adsurl = {http://adsabs.harvard.edu/abs/2012MNRAS.422..926M}
}

@ARTICLE{Harnois12,
   author = {{Harnois-D{\'e}raps}, J. and {Pen}, U.-L. and {Iliev}, I.~T. and 
	{Merz}, H. and {Emberson}, J.~D. and {Desjacques}, V.},
    title = "{High-performance P$^{3}$M N-body code: CUBEP$^{3}$M}",
  journal = {\mnras},
archivePrefix = "arXiv",
   eprint = {1208.5098},
 primaryClass = "astro-ph.CO",
     year = 2013,
    month = nov,
   volume = 436,
    pages = {540-559},
      doi = {10.1093/mnras/stt1591},
   adsurl = {http://adsabs.harvard.edu/abs/2013MNRAS.436..540H}
}

@ARTICLE{2019ApJ...884....1B,
       author = {{Barry}, N. and {Wilensky}, M. and {Trott}, C.~M. and {Pindor}, B. and
         {Beardsley}, A.~P. and {Hazelton}, B.~J. and {Sullivan}, I.~S. and
         {Morales}, M.~F. and {Pober}, J.~C. and {Line}, J. and {Greig}, B. and
         {Byrne}, R. and {Lanman}, A. and {Li}, W. and {Jordan}, C.~H. and
         {Joseph}, R.~C. and {McKinley}, B. and {Rahimi}, M. and {Yoshiura}, S. and
         {Bowman}, J.~D. and {Gaensler}, B.~M. and {Hewitt}, J.~N. and
         {Jacobs}, D.~C. and {Mitchell}, D.~A. and {Udaya Shankar}, N. and
         {Sethi}, S.~K. and {Subrahmanyan}, R. and {Tingay}, S.~J. and
         {Webster}, R.~L. and {Wyithe}, J.~S.~B.},
        title = "{Improving the Epoch of Reionization Power Spectrum Results from Murchison Widefield Array Season 1 Observations}",
      journal = {\apj},
         year = "2019",
        month = "Oct",
       volume = {884},
       number = {1},
          eid = {1},
        pages = {1},
          doi = {10.3847/1538-4357/ab40a8},
archivePrefix = {arXiv},
       eprint = {1909.00561},
 primaryClass = {astro-ph.IM},
       adsurl = {https://ui.adsabs.harvard.edu/abs/2019ApJ...884....1B}
}

@ARTICLE{2012MNRAS.426.3178M,
   author = {{Majumdar}, S. and {Bharadwaj}, S. and {Choudhury}, T.~R.},
    title = "{Constrainingquasar and intergalactic medium properties through bubble detection in redshifted 21-cm maps}",
  journal = {\mnras},
archivePrefix = "arXiv",
   eprint = {1111.6354},
 primaryClass = "astro-ph.CO",
     year = 2012,
    month = nov,
   volume = 426,
    pages = {3178-3194},
      doi = {10.1111/j.1365-2966.2012.21914.x},
   adsurl = {http://adsabs.harvard.edu/abs/2012MNRAS.426.3178M}
}

@ARTICLE{1992PhRvD..46.5236F,
       author = {{Finn}, Lee S.},
        title = "{Detection, measurement, and gravitational radiation}",
      journal = {\prd},
         year = 1992,
        month = Dec,
       volume = {46},
        pages = {5236-5249},
          doi = {10.1103/PhysRevD.46.5236},
archivePrefix = {arXiv},
       eprint = {gr-qc/9209010},
 primaryClass = {gr-qc},
       adsurl = {https://ui.adsabs.harvard.edu/\#abs/1992PhRvD..46.5236F}
}

@ARTICLE{2019MNRAS.488.4271G,
       author = {{Gehlot}, B.~K. and {Mertens}, F.~G. and {Koopmans}, L.~V.~E. and
         {Brentjens}, M.~A. and {Zaroubi}, S. and {Ciardi}, B. and {Ghosh}, A. and
         {Hatef}, M. and {Iliev}, I.~T. and {Jeli{\'c}} and {}, V. and
         {Kooistra}, R. and {Krause}, F. and {Mellema}, G. and {Mevius}, M. and
         {Mitra}, M. and {Offringa}, A.~R. and {Pandey}, V.~N. and
         {Sardarabadi}, A.~M. and {Schaye}, J. and {Silva}, M.~B. and
         {Vedantham}, H.~K. and {Yatawatta}, S.},
        title = "{The first power spectrum limit on the 21-cm signal of neutral hydrogen during the Cosmic Dawn at z = 20-25 from LOFAR}",
      journal = {\mnras},
         year = "2019",
        month = "Sep",
       volume = {488},
       number = {3},
        pages = {4271-4287},
          doi = {10.1093/mnras/stz1937},
archivePrefix = {arXiv},
       eprint = {1809.06661},
 primaryClass = {astro-ph.IM},
       adsurl = {https://ui.adsabs.harvard.edu/abs/2019MNRAS.488.4271G}
}

@ARTICLE{2014PhRvD..90b3019L,
       author = {{Liu}, Adrian and {Parsons}, Aaron R. and {Trott}, Cathryn M.},
        title = "{Epoch of reionization window. II. Statistical methods for foreground wedge reduction}",
      journal = {\prd},
         year = "2014",
        month = "Jul",
       volume = {90},
          eid = {023019},
        pages = {023019},
          doi = {10.1103/PhysRevD.90.023019},
archivePrefix = {arXiv},
       eprint = {1404.4372},
 primaryClass = {astro-ph.CO},
       adsurl = {https://ui.adsabs.harvard.edu/abs/2014PhRvD..90b3019L}
}

@ARTICLE{ghara15b,
   author = {{Ghara}, R. and {Datta}, K.~K. and {Choudhury}, T.~R.},
    title = "{21 cm signal from cosmic dawn - II. Imprints of the light-cone effects}",
  journal = {\mnras},
archivePrefix = "arXiv",
   eprint = {1504.05601},
     year = 2015,
    month = nov,
   volume = 453,
    pages = {3143-3156},
      doi = {10.1093/mnras/stv1855},
   adsurl = {http://adsabs.harvard.edu/abs/2015MNRAS.453.3143G}
}

@ARTICLE{2020ApJ...888...70K,
       author = {{Kern}, Nicholas S. and {Parsons}, Aaron R. and {Dillon}, Joshua S. and {Lanman}, Adam E. and {Liu}, Adrian and {Bull}, Philip and {Ewall-Wice}, Aaron and {Abdurashidova}, Zara and {Aguirre}, James E. and {Alexander}, Paul and {Ali}, Zaki S. and {Balfour}, Yanga and {Beardsley}, Adam P. and {Bernardi}, Gianni and {Bowman}, Judd D. and {Bradley}, Richard F. and {Burba}, Jacob and {Carilli}, Chris L. and {Cheng}, Carina and {DeBoer}, David R. and {Dexter}, Matt and {de Lera Acedo}, Eloy and {Fagnoni}, Nicolas and {Fritz}, Randall and {Furlanetto}, Steve R. and {Glendenning}, Brian and {Gorthi}, Deepthi and {Greig}, Bradley and {Grobbelaar}, Jasper and {Halday}, Ziyaad and {Hazelton}, Bryna J. and {Hewitt}, Jacqueline N. and {Hickish}, Jack and {Jacobs}, Daniel C. and {Julius}, Austin and {Kerrigan}, Joshua and {Kittiwisit}, Piyanat and {Kohn}, Saul A. and {Kolopanis}, Matthew and {La Plante}, Paul and {Lekalake}, Telalo and {MacMahon}, David and {Malan}, Lourence and {Malgas}, Cresshim and {Maree}, Matthys and {Martinot}, Zachary E. and {Matsetela}, Eunice and {Mesinger}, Andrei and {Molewa}, Mathakane and {Morales}, Miguel F. and {Mosiane}, Tshegofalang and {Murray}, Steven G. and {Neben}, Abraham R. and {Parsons}, Aaron R. and {Patra}, Nipanjana and {Pieterse}, Samantha and {Pober}, Jonathan C. and {Razavi-Ghods}, Nima and {Ringuette}, Jon and {Robnett}, James and {Rosie}, Kathryn and {Sims}, Peter and {Smith}, Craig and {Syce}, Angelo and {Thyagarajan}, Nithyanandan and {Williams}, Peter K.~G. and {Zheng}, Haoxuan},
        title = "{Mitigating Internal Instrument Coupling for 21 cm Cosmology. II. A Method Demonstration with the Hydrogen Epoch of Reionization Array}",
      journal = {\apj},
         year = 2020,
        month = jan,
       volume = {888},
       number = {2},
          eid = {70},
        pages = {70},
          doi = {10.3847/1538-4357/ab5e8a},
archivePrefix = {arXiv},
       eprint = {1909.11733},
 primaryClass = {astro-ph.IM},
       adsurl = {https://ui.adsabs.harvard.edu/abs/2020ApJ...888...70K}
}

@ARTICLE{ghara15a,
   author = {{Ghara}, R. and {Choudhury}, T.~R. and {Datta}, K.~K.},
    title = "{21 cm signal from cosmic dawn: imprints of spin temperature fluctuations and peculiar velocities}",
  journal = {\mnras},
archivePrefix = "arXiv",
   eprint = {1406.4157},
     year = 2015,
    month = feb,
   volume = 447,
    pages = {1806-1825},
      doi = {10.1093/mnras/stu2512},
   adsurl = {http://adsabs.harvard.edu/abs/2015MNRAS.447.1806G}
}

@ARTICLE{Gan2022,
       author = {{Gan}, H. and {Koopmans}, L.~V.~E. and {Mertens}, F.~G. and {Mevius}, M. and {Offringa}, A.~R. and {Ciardi}, B. and {Gehlot}, B.~K. and {Ghara}, R. and {Ghosh}, A. and {Giri}, S.~K. and {Iliev}, I.~T. and {Mellema}, G. and {Pandey}, V.~N. and {Zaroubi}, S.},
        title = "{Statistical analysis of the causes of excess variance in the 21 cm signal power spectra obtained with the Low-Frequency Array}",
      journal = {\aap},
         year = 2022,
        month = jul,
       volume = {663},
          eid = {A9},
        pages = {A9},
          doi = {10.1051/0004-6361/202142945},
archivePrefix = {arXiv},
       eprint = {2203.02345},
 primaryClass = {astro-ph.CO},
       adsurl = {https://ui.adsabs.harvard.edu/abs/2022A&A...663A...9G}
}

@ARTICLE{ghara15c,
   author = {{Ghara}, R. and {Choudhury}, T.~R. and {Datta}, K.~K.},
    title = "{21-cm signature of the first sources in the Universe: prospects of detection with SKA}",
  journal = {\mnras},
archivePrefix = "arXiv",
   eprint = {1511.07448},
     year = 2016,
    month = jul,
   volume = 460,
    pages = {827-843},
      doi = {10.1093/mnras/stw953},
   adsurl = {http://adsabs.harvard.edu/abs/2016MNRAS.460..827G}
}

@ARTICLE{ghara16,
   author = {{Ghara}, R. and {Choudhury}, T.~R. and {Datta}, K.~K. and {Choudhuri}, S.
	},
    title = "{Imaging the redshifted 21 cm pattern around the first sources during the cosmic dawn using the SKA}",
  journal = {\mnras},
archivePrefix = "arXiv",
   eprint = {1607.02779},
     year = 2017,
    month = jan,
   volume = 464,
    pages = {2234-2248},
      doi = {10.1093/mnras/stw2494},
   adsurl = {http://adsabs.harvard.edu/abs/2017MNRAS.464.2234G}
}

@ARTICLE{2019JCAP...02..058G,
       author = {{Giri}, Sambit K. and {D'Aloisio}, Anson and {Mellema}, Garrelt and
         {Komatsu}, Eiichiro and {Ghara}, Raghunath and {Majumdar}, Suman},
        title = "{Position-dependent power spectra of the 21-cm signal from the epoch of reionization}",
      journal = {\jcap},
         year = "2019",
        month = "Feb",
       volume = {2019},
       number = {2},
          eid = {058},
        pages = {058},
          doi = {10.1088/1475-7516/2019/02/058},
archivePrefix = {arXiv},
       eprint = {1811.09633},
 primaryClass = {astro-ph.CO},
       adsurl = {https://ui.adsabs.harvard.edu/abs/2019JCAP...02..058G}
}

@ARTICLE{Furlanetto2006,
   author = {{Furlanetto}, S.~R. and {Oh}, S.~P. and {Briggs}, F.~H.},
    title = "{Cosmology at low frequencies: The 21 cm transition and the high-redshift Universe}",
  journal = {\physrep},
   eprint = {astro-ph/0608032},
     year = 2006,
    month = oct,
   volume = 433,
    pages = {181-301},
      doi = {10.1016/j.physrep.2006.08.002},
   adsurl = {http://adsabs.harvard.edu/abs/2006PhR...433..181F}
}

@ARTICLE{Ellis13,
   author = {{Ellis}, R.~S. and {McLure}, R.~J. and {Dunlop}, J.~S. and {Robertson}, B.~E. and 
	{Ono}, Y. and {Schenker}, M.~A. and {Koekemoer}, A. and {Bowler}, R.~A.~A. and 
	{Ouchi}, M. and {Rogers}, A.~B. and {Curtis-Lake}, E. and {Schneider}, E. and 
	{Charlot}, S. and {Stark}, D.~P. and {Furlanetto}, S.~R. and 
	{Cirasuolo}, M.},
    title = "{The Abundance of Star-forming Galaxies in the Redshift Range 8.5-12: New Results from the 2012 Hubble Ultra Deep Field Campaign}",
  journal = {\apjl},
archivePrefix = "arXiv",
   eprint = {1211.6804},
     year = 2013,
    month = jan,
   volume = 763,
      eid = {L7},
    pages = {L7},
      doi = {10.1088/2041-8205/763/1/L7},
   adsurl = {http://adsabs.harvard.edu/abs/2013ApJ...763L...7E}
}

@ARTICLE{Fan06b,
   author = {{Fan}, X. and {Strauss}, M.~A. and {Becker}, R.~H. and {White}, R.~L. and 
	{Gunn}, J.~E. and {Knapp}, G.~R. and {Richards}, G.~T. and {Schneider}, D.~P. and 
	{Brinkmann}, J. and {Fukugita}, M.},
    title = "{Constraining the Evolution of the Ionizing Background and the Epoch of Reionization with z\~{}6 Quasars. II. A Sample of 19 Quasars}",
  journal = {\aj},
   eprint = {astro-ph/0512082},
     year = 2006,
    month = jul,
   volume = 132,
    pages = {117-136},
      doi = {10.1086/504836},
   adsurl = {http://adsabs.harvard.edu/abs/2006AJ....132..117F}
}

@ARTICLE{Planck2013,
   author = {{Planck Collaboration} and {Ade}, P.~A.~R. and {Aghanim}, N. and 
	{Armitage-Caplan}, C. and {Arnaud}, M. and {Ashdown}, M. and 
	{Atrio-Barandela}, F. and {Aumont}, J. and {Baccigalupi}, C. and 
	{Banday}, A.~J. and et al.},
    title = "{Planck 2013 results. XVI. Cosmological parameters}",
  journal = {\aap},
archivePrefix = "arXiv",
   eprint = {1303.5076},
     year = 2014,
    month = nov,
   volume = 571,
      eid = {A16},
    pages = {A16},
      doi = {10.1051/0004-6361/201321591},
   adsurl = {http://adsabs.harvard.edu/abs/2014A%26A...571A..16P}
}

@ARTICLE{Venemans15,
   author = {{Venemans}, B.~P. and {Ba{\~n}ados}, E. and {Decarli}, R. and 
	{Farina}, E.~P. and {Walter}, F. and {Chambers}, K.~C. and {Fan}, X. and 
	{Rix}, H.-W. and {Schlafly}, E. and {McMahon}, R.~G. and {Simcoe}, R. and 
	{Stern}, D. and {Burgett}, W.~S. and {Draper}, P.~W. and {Flewelling}, H. and 
	{Hodapp}, K.~W. and {Kaiser}, N. and {Magnier}, E.~A. and {Metcalfe}, N. and 
	{Morgan}, J.~S. and {Price}, P.~A. and {Tonry}, J.~L. and {Waters}, C. and 
	{AlSayyad}, Y. and {Banerji}, M. and {Chen}, S.~S. and {Gonz{\'a}lez-Solares}, E.~A. and 
	{Greiner}, J. and {Mazzucchelli}, C. and {McGreer}, I. and {Miller}, D.~R. and 
	{Reed}, S. and {Sullivan}, P.~W.},
    title = "{The Identification of Z-dropouts in Pan-STARRS1: Three Quasars at 6.5< z< 6.7}",
  journal = {\apjl},
archivePrefix = "arXiv",
   eprint = {1502.01927},
     year = 2015,
    month = mar,
   volume = 801,
      eid = {L11},
    pages = {L11},
      doi = {10.1088/2041-8205/801/1/L11},
   adsurl = {http://adsabs.harvard.edu/abs/2015ApJ...801L..11V}
}

@ARTICLE{madau1997,
   author = {{Madau}, P. and {Meiksin}, A. and {Rees}, M.~J.},
    title = "{21 Centimeter Tomography of the Intergalactic Medium at High Redshift}",
  journal = {\apj},
   eprint = {astro-ph/9608010},
     year = 1997,
    month = feb,
   volume = 475,
    pages = {429-444},
   adsurl = {http://adsabs.harvard.edu/abs/1997ApJ...475..429M}
}

@ARTICLE{2018MNRAS.475.1213S,
       author = {{Schmit}, C.~J. and {Pritchard}, J.~R.},
        title = "{Emulation of reionization simulations for Bayesian inference of astrophysics parameters using neural networks}",
      journal = {\mnras},
         year = "2018",
        month = "Mar",
       volume = {475},
        pages = {1213-1223},
          doi = {10.1093/mnras/stx3292},
archivePrefix = {arXiv},
       eprint = {1708.00011},
 primaryClass = {astro-ph.CO},
       adsurl = {https://ui.adsabs.harvard.edu/\#abs/2018MNRAS.475.1213S}
}

@ARTICLE{Harikane24,
       author = {{Harikane}, Yuichi and {Nakajima}, Kimihiko and {Ouchi}, Masami and {Umeda}, Hiroya and {Isobe}, Yuki and {Ono}, Yoshiaki and {Xu}, Yi and {Zhang}, Yechi},
        title = "{Pure Spectroscopic Constraints on UV Luminosity Functions and Cosmic Star Formation History from 25 Galaxies at z $_{spec}$ = 8.61-13.20 Confirmed with JWST/NIRSpec}",
      journal = {\apj},
         year = 2024,
        month = jan,
       volume = {960},
       number = {1},
          eid = {56},
        pages = {56},
          doi = {10.3847/1538-4357/ad0b7e},
archivePrefix = {arXiv},
       eprint = {2304.06658},
 primaryClass = {astro-ph.GA},
       adsurl = {https://ui.adsabs.harvard.edu/abs/2024ApJ...960...56H}
}

@ARTICLE{Bouwens23,
       author = {{Bouwens}, Rychard and {Illingworth}, Garth and {Oesch}, Pascal and {Stefanon}, Mauro and {Naidu}, Rohan and {van Leeuwen}, Ivana and {Magee}, Dan},
        title = "{UV luminosity density results at z > 8 from the first JWST/NIRCam fields: limitations of early data sets and the need for spectroscopy}",
      journal = {\mnras},
         year = 2023,
        month = jul,
       volume = {523},
       number = {1},
        pages = {1009-1035},
          doi = {10.1093/mnras/stad1014},
archivePrefix = {arXiv},
       eprint = {2212.06683},
 primaryClass = {astro-ph.CO},
       adsurl = {https://ui.adsabs.harvard.edu/abs/2023MNRAS.523.1009B}
}

@ARTICLE{Mortlock11,
   author = {{Mortlock}, D.~J. and {Warren}, S.~J. and {Venemans}, B.~P. and 
	{Patel}, M. and {Hewett}, P.~C. and {McMahon}, R.~G. and {Simpson}, C. and 
	{Theuns}, T. and {Gonz{\'a}les-Solares}, E.~A. and {Adamson}, A. and 
	{Dye}, S. and {Hambly}, N.~C. and {Hirst}, P. and {Irwin}, M.~J. and 
	{Kuiper}, E. and {Lawrence}, A. and {R{\"o}ttgering}, H.~J.~A.
	},
    title = "{A luminous quasar at a redshift of z = 7.085}",
  journal = {\nat},
archivePrefix = "arXiv",
   eprint = {1106.6088},
 primaryClass = "astro-ph.CO",
     year = 2011,
    month = jun,
   volume = 474,
    pages = {616-619},
      doi = {10.1038/nature10159},
   adsurl = {http://adsabs.harvard.edu/abs/2011Natur.474..616M}
}

@ARTICLE{2025MNRAS.tmp.1333M,
       author = {{Munshi}, S. and {Mertens}, F.~G. and {Chege}, J.~K. and {Koopmans}, L.~V.~E. and {Offringa}, A.~R. and {Semelin}, B. and {Barkana}, R. and {Dhandha}, J. and {Fialkov}, A. and {M{\'e}riot}, R. and {Sikder}, S. and {Bracco}, A. and {Brackenhoff}, S.~A. and {Ceccotti}, E. and {Ghara}, R. and {Ghosh}, S. and {Hothi}, I. and {Mevius}, M. and {Ocvirk}, P. and {Shaw}, A.~K. and {Yatawatta}, S. and {Zarka}, P.},
        title = "{Improved upper limits on the 21-cm signal power spectrum at z = 17.0 and z = 20.3 from an optimal field observed with NenuFAR}",
      journal = {\mnras},
         year = 2025,
        month = aug,
          doi = {10.1093/mnras/staf1386},
archivePrefix = {arXiv},
       eprint = {2507.10533},
 primaryClass = {astro-ph.CO},
       adsurl = {https://ui.adsabs.harvard.edu/abs/2025MNRAS.tmp.1333M}
}

@ARTICLE{2024JCAP...10..003N,
       author = {{Noble}, Leon and {Kamran}, Mohd and {Majumdar}, Suman and {Murmu}, Chandra Shekhar and {Ghara}, Raghunath and {Mellema}, Garrelt and {Iliev}, Ilian T. and {Pritchard}, Jonathan R.},
        title = "{Impact of the Epoch of Reionization sources on the 21-cm bispectrum}",
      journal = {\jcap},
         year = 2024,
        month = oct,
       volume = {2024},
       number = {10},
          eid = {003},
        pages = {003},
          doi = {10.1088/1475-7516/2024/10/003},
archivePrefix = {arXiv},
       eprint = {2406.03118},
 primaryClass = {astro-ph.CO},
       adsurl = {https://ui.adsabs.harvard.edu/abs/2024JCAP...10..003N}
}

@ARTICLE{2019MNRAS.487.2785I,
       author = {{Islam}, Nazma and {Ghara}, Raghunath and {Paul}, Biswajit and
         {Choudhury}, T. Roy and {Nath}, Biman B.},
        title = "{Cosmological implications of the composite spectra of galactic X-ray binaries constructed using MAXI data}",
      journal = {\mnras},
         year = "2019",
        month = "Aug",
       volume = {487},
       number = {2},
        pages = {2785-2796},
          doi = {10.1093/mnras/stz1446},
archivePrefix = {arXiv},
       eprint = {1905.10386},
 primaryClass = {astro-ph.HE},
       adsurl = {https://ui.adsabs.harvard.edu/abs/2019MNRAS.487.2785I}
}

@ARTICLE{Ross2019,
       author = {{Ross}, Hannah E. and {Dixon}, Keri L. and {Ghara}, Raghunath and {Iliev}, Ilian T. and {Mellema}, Garrelt},
        title = "{Evaluating the QSO contribution to the 21-cm signal from the Cosmic Dawn}",
      journal = {\mnras},
         year = 2019,
        month = jul,
       volume = {487},
       number = {1},
        pages = {1101-1119},
          doi = {10.1093/mnras/stz1220},
       adsurl = {https://ui.adsabs.harvard.edu/abs/2019MNRAS.487.1101R}
}

@ARTICLE{2018MNRAS.478.3640M,
       author = {{Mertens}, F.~G. and {Ghosh}, A. and {Koopmans}, L.~V.~E.},
        title = "{Statistical 21-cm signal separation via Gaussian Process Regression analysis}",
      journal = {\mnras},
         year = "2018",
        month = "Aug",
       volume = {478},
       number = {3},
        pages = {3640-3652},
          doi = {10.1093/mnras/sty1207},
archivePrefix = {arXiv},
       eprint = {1711.10834},
 primaryClass = {astro-ph.CO},
       adsurl = {https://ui.adsabs.harvard.edu/abs/2018MNRAS.478.3640M}
}

@ARTICLE{2020MNRAS.496..739G,
       author = {{Ghara}, Raghunath and {Choudhury}, T. Roy},
        title = "{Bayesian approach to constraining the properties of ionized bubbles during reionization}",
      journal = {\mnras},
         year = 2020,
        month = jun,
       volume = {496},
       number = {1},
        pages = {739-753},
          doi = {10.1093/mnras/staa1599},
archivePrefix = {arXiv},
       eprint = {1909.12317},
 primaryClass = {astro-ph.CO},
       adsurl = {https://ui.adsabs.harvard.edu/abs/2020MNRAS.496..739G}
}

@ARTICLE{2021JCAP...05..026K,
       author = {{Kapahtia}, Akanksha and {Chingangbam}, Pravabati and {Ghara}, Raghunath and {Appleby}, Stephen and {Choudhury}, Tirthankar Roy},
        title = "{Prospects of constraining reionization model parameters using Minkowski tensors and Betti numbers}",
      journal = {\jcap},
         year = 2021,
        month = may,
       volume = {2021},
       number = {5},
          eid = {026},
        pages = {026},
          doi = {10.1088/1475-7516/2021/05/026},
archivePrefix = {arXiv},
       eprint = {2101.03962},
 primaryClass = {astro-ph.CO},
       adsurl = {https://ui.adsabs.harvard.edu/abs/2021JCAP...05..026K}
}

@ARTICLE{2022JCAP...11..001K,
       author = {{Kamran}, Mohd and {Ghara}, Raghunath and {Majumdar}, Suman and {Mellema}, Garrelt and {Bharadwaj}, Somnath and {Pritchard}, Jonathan R. and {Mondal}, Rajesh and {Iliev}, Ilian T.},
        title = "{Redshifted 21-cm bispectrum: impact of the source models on the signal and the IGM physics from the Cosmic Dawn}",
      journal = {\jcap},
         year = 2022,
        month = nov,
       volume = {2022},
       number = {11},
          eid = {001},
        pages = {001},
          doi = {10.1088/1475-7516/2022/11/001},
archivePrefix = {arXiv},
       eprint = {2207.09128},
 primaryClass = {astro-ph.CO},
       adsurl = {https://ui.adsabs.harvard.edu/abs/2022JCAP...11..001K}
}

@ARTICLE{2021MNRAS.506.3717R,
       author = {{Ross}, Hannah E. and {Giri}, Sambit K. and {Mellema}, Garrelt and {Dixon}, Keri L. and {Ghara}, Raghunath and {Iliev}, Ilian T.},
        title = "{Redshift-space distortions in simulations of the 21-cm signal from the cosmic dawn}",
      journal = {\mnras},
         year = 2021,
        month = sep,
       volume = {506},
       number = {3},
        pages = {3717-3733},
          doi = {10.1093/mnras/stab1822},
archivePrefix = {arXiv},
       eprint = {2011.03558},
 primaryClass = {astro-ph.CO},
       adsurl = {https://ui.adsabs.harvard.edu/abs/2021MNRAS.506.3717R}
}

@ARTICLE{2017MNRAS.468.3869S,
   author = {{Shimabukuro}, H. and {Semelin}, B.},
    title = "{Analysing the 21 cm signal from the epoch of reionization with artificial neural networks}",
  journal = {\mnras},
archivePrefix = "arXiv",
   eprint = {1701.07026},
     year = 2017,
    month = jul,
   volume = 468,
    pages = {3869-3877},
      doi = {10.1093/mnras/stx734},
   adsurl = {http://adsabs.harvard.edu/abs/2017MNRAS.468.3869S}
}

@ARTICLE{2020arXiv200603203G,
       author = {{Greig}, Bradley and {Mesinger}, Andrei and {Koopmans}, L{\'e}on V.~E. and {Ciardi}, Benedetta and {Mellema}, Garrelt and {Zaroubi}, Saleem and {Giri}, Sambit K. and {Ghara}, Raghunath and {Ghosh}, Abhik and {Iliev}, Ilian T. and {Mertens}, Florent G. and {Mondal}, Rajesh and {Offringa}, Andr{\'e} R. and {Pandey}, Vishambhar N.},
        title = "{Interpreting LOFAR 21-cm signal upper limits at z = 9.1 in the context of high-z galaxy and reionization observations}",
      journal = {\mnras},
         year = 2021,
        month = jan,
       volume = {501},
       number = {1},
        pages = {1-13},
          doi = {10.1093/mnras/staa3593},
archivePrefix = {arXiv},
       eprint = {2006.03203},
 primaryClass = {astro-ph.CO},
       adsurl = {https://ui.adsabs.harvard.edu/abs/2021MNRAS.501....1G}
}

@ARTICLE{2023A&A...669A..20G,
       author = {{Gan}, H. and {Mertens}, F.~G. and {Koopmans}, L.~V.~E. and {Offringa}, A.~R. and {Mevius}, M. and {Pandey}, V.~N. and {Brackenhoff}, S.~A. and {Ceccotti}, E. and {Ciardi}, B. and {Gehlot}, B.~K. and {Ghara}, R. and {Giri}, S.~K. and {Iliev}, I.~T. and {Munshi}, S.},
        title = "{Assessing the impact of two independent direction-dependent calibration algorithms on the LOFAR 21 cm signal power spectrum. And applications to an observation of a field flanking the north celestial pole}",
      journal = {\aap},
         year = 2023,
        month = jan,
       volume = {669},
          eid = {A20},
        pages = {A20},
          doi = {10.1051/0004-6361/202244316},
       adsurl = {https://ui.adsabs.harvard.edu/abs/2023A&A...669A..20G}
}

@ARTICLE{2020MNRAS.493.4728G,
       author = {{Ghara}, R. and {Giri}, S.~K. and {Mellema}, G. and {Ciardi}, B. and
         {Zaroubi}, S. and {Iliev}, I.~T. and {Koopmans}, L.~V.~E. and
         {Chapman}, E. and {Gazagnes}, S. and {Gehlot}, B.~K. and {Ghosh}, A. and
         {Jeli{\'c}}, V. and {Mertens}, F.~G. and {Mondal}, R. and {Schaye}, J. and
         {Silva}, M.~B. and {Asad}, K.~M.~B. and {Kooistra}, R. and
         {Mevius}, M. and {Offringa}, A.~R. and {Pandey}, V.~N. and
         {Yatawatta}, S.},
        title = "{Constraining the intergalactic medium at z {\ensuremath{\approx}} 9.1 using LOFAR Epoch of Reionization observations}",
      journal = {\mnras},
         year = 2020,
        month = apr,
       volume = {493},
       number = {4},
        pages = {4728-4747},
          doi = {10.1093/mnras/staa487},
archivePrefix = {arXiv},
       eprint = {2002.07195},
 primaryClass = {astro-ph.CO},
       adsurl = {https://ui.adsabs.harvard.edu/abs/2020MNRAS.493.4728G}
}

@ARTICLE{2020MNRAS.493.1662M,
       author = {{Mertens}, F.~G. and {Mevius}, M. and {Koopmans}, L.~V.~E. and
         {Offringa}, A.~R. and {Mellema}, G. and {Zaroubi}, S. and
         {Brentjens}, M.~A. and {Gan}, H. and {Gehlot}, B.~K. and {Pand
        ey}, V.~N. and {Sardarabadi}, A.~M. and {Vedantham}, H.~K. and
         {Yatawatta}, S. and {Asad}, K.~M.~B. and {Ciardi}, B. and
         {Chapman}, E. and {Gazagnes}, S. and {Ghara}, R. and {Ghosh}, A. and
         {Giri}, S.~K. and {Iliev}, I.~T. and {Jeli{\'c}}, V. and
         {Kooistra}, R. and {Mondal}, R. and {Schaye}, J. and {Silva}, M.~B.},
        title = "{Improved upper limits on the 21 cm signal power spectrum of neutral hydrogen at z {\ensuremath{\approx}} 9.1 from LOFAR}",
      journal = {\mnras},
         year = 2020,
        month = apr,
       volume = {493},
       number = {2},
        pages = {1662-1685},
          doi = {10.1093/mnras/staa327},
archivePrefix = {arXiv},
       eprint = {2002.07196},
 primaryClass = {astro-ph.CO},
       adsurl = {https://ui.adsabs.harvard.edu/abs/2020MNRAS.493.1662M}
}

@ARTICLE{lofar2025,
       author = {{Mertens}, F.~G. and {Mevius}, M. and {Koopmans}, L.~V.~E. and {Offringa}, A.~R. and {Zaroubi}, S. and {Acharya}, A. and {Brackenhoff}, S.~A. and {Ceccotti}, E. and {Chapman}, E. and {Chege}, K. and {Ciardi}, B. and {Ghara}, R. and {Ghosh}, S. and {Giri}, S.~K. and {Hothi}, I. and {H{\"o}fer}, C. and {Iliev}, I.~T. and {Jeli{\'c}}, V. and {Ma}, Q. and {Mellema}, G. and {Munshi}, S. and {Pandey}, V.~N. and {Yatawatta}, S.},
        title = "{Deeper multi-redshift upper limits on the epoch of reionisation 21 cm signal power spectrum from LOFAR between z = 8.3 and z = 10.1}",
      journal = {\aap},
         year = 2025,
        month = jun,
       volume = {698},
          eid = {A186},
        pages = {A186},
          doi = {10.1051/0004-6361/202554158},
archivePrefix = {arXiv},
       eprint = {2503.05576},
 primaryClass = {astro-ph.CO},
       adsurl = {https://ui.adsabs.harvard.edu/abs/2025A&A...698A.186M}
}

@ARTICLE{2025A&A...697A.203M,
       author = {{Munshi}, S. and {Mertens}, F.~G. and {Koopmans}, L.~V.~E. and {Mevius}, M. and {Offringa}, A.~R. and {Semelin}, B. and {Viou}, C. and {Bracco}, A. and {Brackenhoff}, S.~A. and {Ceccotti}, E. and {Chege}, J.~K. and {Fialkov}, A. and {Gao}, L.~Y. and {Ghara}, R. and {Ghosh}, S. and {Shaw}, A.~K. and {Zarka}, P. and {Zaroubi}, S. and {Cecconi}, B. and {Corbel}, S. and {Girard}, J.~N. and {Grie{\ss}meier}, J.~M. and {Konovalenko}, O. and {Loh}, A. and {Tokarsky}, P. and {Ulyanov}, O. and {Zakharenko}, V.},
        title = "{Near-field imaging of local interference in radio interferometric data: Impact on the redshifted 21 cm power spectrum}",
      journal = {\aap},
         year = 2025,
        month = may,
       volume = {697},
          eid = {A203},
        pages = {A203},
          doi = {10.1051/0004-6361/202554763},
archivePrefix = {arXiv},
       eprint = {2503.21728},
 primaryClass = {astro-ph.IM},
       adsurl = {https://ui.adsabs.harvard.edu/abs/2025A&A...697A.203M}
}

@ARTICLE{2020MNRAS.498.4178M,
       author = {{Mondal}, R. and {Fialkov}, A. and {Fling}, C. and {Iliev}, I.~T. and {Barkana}, R. and {Ciardi}, B. and {Mellema}, G. and {Zaroubi}, S. and {Koopmans}, L.~V.~E. and {Mertens}, F.~G. and {Gehlot}, B.~K. and {Ghara}, R. and {Ghosh}, A. and {Giri}, S.~K. and {Offringa}, A. and {Pandey}, V.~N.},
        title = "{Tight constraints on the excess radio background at z = 9.1 from LOFAR}",
      journal = {\mnras},
         year = 2020,
        month = nov,
       volume = {498},
       number = {3},
        pages = {4178-4191},
          doi = {10.1093/mnras/staa2422},
archivePrefix = {arXiv},
       eprint = {2004.00678},
 primaryClass = {astro-ph.CO},
       adsurl = {https://ui.adsabs.harvard.edu/abs/2020MNRAS.498.4178M}
}

@ARTICLE{2015MNRAS.449.4246G,
   author = {{Greig}, B. and {Mesinger}, A.},
    title = "{21CMMC: an MCMC analysis tool enabling astrophysical parameter studies of the cosmic 21 cm signal}",
  journal = {\mnras},
archivePrefix = "arXiv",
   eprint = {1501.06576},
     year = 2015,
    month = jun,
   volume = 449,
    pages = {4246-4263},
      doi = {10.1093/mnras/stv571},
   adsurl = {http://adsabs.harvard.edu/abs/2015MNRAS.449.4246G}
}

@ARTICLE{2015aska.confE..10M,
   author = {{Mellema}, G. and {Koopmans}, L. and {Shukla}, H. and {Datta}, K.~K. and 
	{Mesinger}, A. and {Majumdar}, S.},
    title = "{HI tomographic imaging of the Cosmic Dawn and Epoch of Reionization with SKA}",
  journal = {Advancing Astrophysics with the Square Kilometre Array (AASKA14)},
archivePrefix = "arXiv",
   eprint = {1501.04203},
     year = 2015,
      eid = {10},
    pages = {10},
   adsurl = {http://adsabs.harvard.edu/abs/2015aska.confE..10M}
}

@ARTICLE{2019ApJ...883..133K,
       author = {{Kolopanis}, Matthew and {Jacobs}, Daniel C. and {Cheng}, Carina and
         {Parsons}, Aaron R. and {Kohn}, Saul A. and {Pober}, Jonathan C. and
         {Aguirre}, James E. and {Ali}, Zaki S. and {Bernardi}, Gianni and
         {Bradley}, Richard F. and {Carilli}, Chris L. and {DeBoer}, David R. and
         {Dexter}, Matthew R. and {Dillon}, Joshua S. and {Kerrigan}, Joshua and
         {Klima}, Pat and {Liu}, Adrian and {MacMahon}, David H.~E. and
         {Moore}, David F. and {Thyagarajan}, Nithyanandan and
         {Nunhokee}, Chuneeta D. and {Walbrugh}, William P. and {Walker}, Andre},
        title = "{A Simplified, Lossless Reanalysis of PAPER-64}",
      journal = {\apj},
         year = "2019",
        month = "Oct",
       volume = {883},
       number = {2},
          eid = {133},
        pages = {133},
          doi = {10.3847/1538-4357/ab3e3a},
archivePrefix = {arXiv},
       eprint = {1909.02085},
 primaryClass = {astro-ph.CO},
       adsurl = {https://ui.adsabs.harvard.edu/abs/2019ApJ...883..133K}
}

@ARTICLE{2018ApJ...868...26C,
       author = {{Cheng}, Carina and {Parsons}, Aaron R. and {Kolopanis}, Matthew and
         {Jacobs}, Daniel C. and {Liu}, Adrian and {Kohn}, Saul A. and
         {Aguirre}, James E. and {Pober}, Jonathan C. and {Ali}, Zaki S. and
         {Bernardi}, Gianni and {Bradley}, Richard F. and {Carilli}, Chris L. and
         {DeBoer}, David R. and {Dexter}, Matthew R. and {Dillon}, Joshua S. and
         {Klima}, Pat and {MacMahon}, David H.~E. and {Moore}, David F. and
         {Nunhokee}, Chuneeta D. and {Walbrugh}, William P. and {Walker}, Andre},
        title = "{Characterizing Signal Loss in the 21 cm Reionization Power Spectrum: A Revised Study of PAPER-64}",
      journal = {\apj},
         year = "2018",
        month = "Nov",
       volume = {868},
       number = {1},
          eid = {26},
        pages = {26},
          doi = {10.3847/1538-4357/aae833},
archivePrefix = {arXiv},
       eprint = {1810.05175},
 primaryClass = {astro-ph.IM},
       adsurl = {https://ui.adsabs.harvard.edu/abs/2018ApJ...868...26C}
}

@ARTICLE{2019ApJ...884..105K,
       author = {{Kern}, Nicholas S. and {Parsons}, Aaron R. and {Dillon}, Joshua S. and
         {Lanman}, Adam E. and {Fagnoni}, Nicolas and {de Lera Acedo}, Eloy},
        title = "{Mitigating Internal Instrument Coupling for 21 cm Cosmology. I. Temporal and Spectral Modeling in Simulations}",
      journal = {\apj},
         year = "2019",
        month = "Oct",
       volume = {884},
       number = {2},
          eid = {105},
        pages = {105},
          doi = {10.3847/1538-4357/ab3e73},
       adsurl = {https://ui.adsabs.harvard.edu/abs/2019ApJ...884..105K}
}

@ARTICLE{2021A&A...652A..37E,
       author = {{Edler}, H.~W. and {de Gasperin}, F. and {Rafferty}, D.},
        title = "{Investigating ionospheric calibration for LOFAR 2.0 with simulated observations}",
      journal = {\aap},
         year = 2021,
        month = aug,
       volume = {652},
          eid = {A37},
        pages = {A37},
          doi = {10.1051/0004-6361/202140465},
archivePrefix = {arXiv},
       eprint = {2105.04636},
 primaryClass = {astro-ph.IM},
       adsurl = {https://ui.adsabs.harvard.edu/abs/2021A&A...652A..37E}
}

@article{Wilensky_2019,
   title={Absolving the SSINS of Precision Interferometric Radio Data: A New Technique for Mitigating Faint Radio Frequency Interference},
   volume={131},
   ISSN={1538-3873},
   url={http://dx.doi.org/10.1088/1538-3873/ab3cad},
   DOI={10.1088/1538-3873/ab3cad},
   number={1005},
   journal={Publications of the Astronomical Society of the Pacific},
   publisher={IOP Publishing},
   author={Wilensky, Michael J. and Morales, Miguel F. and Hazelton, Bryna J. and Barry, Nichole and Byrne, Ruby and Roy, Sumit},
   year={2019},
   month=oct, pages={114507} }

@ARTICLE{Patil2016,
       author = {{Patil}, Ajinkya H. and {Yatawatta}, Sarod and {Zaroubi}, Saleem and {Koopmans}, L{\'e}on V.~E. and {de Bruyn}, A.~G. and {Jeli{\'c}}, Vibor and {Ciardi}, Benedetta and {Iliev}, Ilian T. and {Mevius}, Maaijke and {Pandey}, Vishambhar N. and {Gehlot}, Bharat K.},
        title = "{Systematic biases in low-frequency radio interferometric data due to calibration: the LOFAR-EoR case}",
      journal = {\mnras},
         year = 2016,
        month = dec,
       volume = {463},
       number = {4},
        pages = {4317-4330},
          doi = {10.1093/mnras/stw2277},
archivePrefix = {arXiv},
       eprint = {1605.07619},
 primaryClass = {astro-ph.IM},
       adsurl = {https://ui.adsabs.harvard.edu/abs/2016MNRAS.463.4317P}
}

@ARTICLE{Ewall2017,
       author = {{Ewall-Wice}, Aaron and {Dillon}, Joshua S. and {Liu}, Adrian and {Hewitt}, Jacqueline},
        title = "{The impact of modelling errors on interferometer calibration for 21 cm power spectra}",
      journal = {\mnras},
         year = 2017,
        month = sep,
       volume = {470},
       number = {2},
        pages = {1849-1870},
          doi = {10.1093/mnras/stx1221},
archivePrefix = {arXiv},
       eprint = {1610.02689},
 primaryClass = {astro-ph.CO},
       adsurl = {https://ui.adsabs.harvard.edu/abs/2017MNRAS.470.1849E}
}

@ARTICLE{offringa2019,
       author = {{Offringa}, A.~R. and {Mertens}, F. and {Koopmans}, L.~V.~E.},
        title = "{The impact of interference excision on 21-cm epoch of reionization power spectrum analyses}",
      journal = {\mnras},
         year = 2019,
        month = apr,
       volume = {484},
       number = {2},
        pages = {2866-2875},
          doi = {10.1093/mnras/stz175},
archivePrefix = {arXiv},
       eprint = {1901.04752},
 primaryClass = {astro-ph.IM},
       adsurl = {https://ui.adsabs.harvard.edu/abs/2019MNRAS.484.2866O}
}

@ARTICLE{2016RaSc...51..927M,
       author = {{Mevius}, M. and {van der Tol}, S. and {Pandey}, V.~N. and
         {Vedantham}, H.~K. and {Brentjens}, M.~A. and {de Bruyn}, A.~G. and
         {Abdalla}, F.~B. and {Asad}, K.~M.~B. and {Bregman}, J.~D. and
         {Brouw}, W.~N. and {Bus}, S. and {Chapman}, E. and {Ciardi}, B. and
         {Fernandez}, E.~R. and {Ghosh}, A. and {Harker}, G. and {Iliev}, I.~T. and
         {Jeli{\'c}}, V. and {Kazemi}, S. and {Koopmans}, L.~V.~E. and
         {Noordam}, J.~E. and {Offringa}, A.~R. and {Patil}, A.~H. and
         {van Weeren}, R.~J. and {Wijnholds}, S. and {Yatawatta}, S. and
         {Zaroubi}, S.},
        title = "{Probing ionospheric structures using the LOFAR radio telescope}",
      journal = {Radio Science},
         year = "2016",
        month = "Jul",
       volume = {51},
       number = {7},
        pages = {927-941},
          doi = {10.1002/2016RS006028},
archivePrefix = {arXiv},
       eprint = {1606.04683},
 primaryClass = {astro-ph.IM},
       adsurl = {https://ui.adsabs.harvard.edu/abs/2016RaSc...51..927M}
}

@ARTICLE{2018MNRAS.475..438R,
       author = {{Raut}, Dinesh and {Choudhury}, Tirthankar Roy and {Ghara}, Raghunath},
        title = "{Measuring the reionization 21 cm fluctuations using clustering wedges}",
      journal = {\mnras},
         year = 2018,
        month = mar,
       volume = {475},
       number = {1},
        pages = {438-447},
          doi = {10.1093/mnras/stx3190},
archivePrefix = {arXiv},
       eprint = {1708.02824},
 primaryClass = {astro-ph.CO},
       adsurl = {https://ui.adsabs.harvard.edu/abs/2018MNRAS.475..438R}
}

@ARTICLE{2016MNRAS.461.3135B,
       author = {{Barry}, N. and {Hazelton}, B. and {Sullivan}, I. and {Morales}, M.~F. and
         {Pober}, J.~C.},
        title = "{Calibration requirements for detecting the 21 cm epoch of reionization power spectrum and implications for the SKA}",
      journal = {\mnras},
         year = "2016",
        month = "Sep",
       volume = {461},
       number = {3},
        pages = {3135-3144},
          doi = {10.1093/mnras/stw1380},
archivePrefix = {arXiv},
       eprint = {1603.00607},
 primaryClass = {astro-ph.IM},
       adsurl = {https://ui.adsabs.harvard.edu/abs/2016MNRAS.461.3135B}
}

@article{Abdurashidova_2023,
doi = {10.3847/1538-4357/acaf50},
url = {https://dx.doi.org/10.3847/1538-4357/acaf50},
year = {2023},
month = {mar},
publisher = {The American Astronomical Society},
volume = {945},
number = {2},
pages = {124},
author = {The HERA Collaboration: Zara Abdurashidova and Tyrone Adams and James E. Aguirre and Paul Alexander and Zaki S. Ali and Rushelle Baartman and Yanga Balfour and Rennan Barkana and Adam P. Beardsley and Gianni Bernardi and Tashalee S. Billings and Judd D. Bowman and Richard F. Bradley and Daniela Breitman and Philip Bull and Jacob Burba and Steve Carey and Chris L. Carilli and Carina Cheng and Samir Choudhuri and David R. DeBoer and Eloy de Lera Acedo and Matt Dexter and Joshua S. Dillon and John Ely and Aaron Ewall-Wice and Nicolas Fagnoni and Anastasia Fialkov and Randall Fritz and Steven R. Furlanetto and Kingsley Gale-Sides and Hugh Garsden and Brian Glendenning and Adélie Gorce and Deepthi Gorthi and Bradley Greig and Jasper Grobbelaar and Ziyaad Halday and Bryna J. Hazelton and Stefan Heimersheim and Jacqueline N. Hewitt and Jack Hickish and Daniel C. Jacobs and Austin Julius and Nicholas S. Kern and Joshua Kerrigan and Piyanat Kittiwisit and Saul A. Kohn and Matthew Kolopanis and Adam Lanman and Paul La Plante and David Lewis and Adrian Liu and Anita Loots and Yin-Zhe Ma and David H. E. MacMahon and Lourence Malan and Keith Malgas and Cresshim Malgas and Matthys Maree and Bradley Marero and Zachary E. Martinot and Lisa McBride and Andrei Mesinger and Jordan Mirocha and Mathakane Molewa and Miguel F. Morales and Tshegofalang Mosiane and Julian B. Muñoz and Steven G. Murray and Vighnesh Nagpal and Abraham R. Neben and Bojan Nikolic and Chuneeta D. Nunhokee and Hans Nuwegeld and Aaron R. Parsons and Robert Pascua and Nipanjana Patra and Samantha Pieterse and Yuxiang Qin and Nima Razavi-Ghods and James Robnett and Kathryn Rosie and Mario G. Santos and Peter Sims and Saurabh Singh and Craig Smith and Hilton Swarts and Jianrong Tan and Nithyanandan Thyagarajan and Michael J. Wilensky and Peter K. G. Williams and Pieter van Wyngaarden and Haoxuan Zheng},
title = {Improved Constraints on the 21 cm EoR Power Spectrum and the X-Ray Heating of the IGM with HERA Phase I Observations},
journal = {The Astrophysical Journal}
}

@ARTICLE{2023MNRAS.522.2188S,
       author = {{Shaw}, Abinash Kumar and {Ghara}, Raghunath and {Zaroubi}, Saleem and {Mondal}, Rajesh and {Mellema}, Garrelt and {Mertens}, Florent and {Koopmans}, L{\'e}on V.~E. and {Semelin}, Beno{\^\i}t},
        title = "{Studying the multifrequency angular power spectrum of the cosmic dawn 21-cm signal}",
      journal = {\mnras},
         year = 2023,
        month = jun,
       volume = {522},
       number = {2},
        pages = {2188-2206},
          doi = {10.1093/mnras/stad1114},
archivePrefix = {arXiv},
       eprint = {2302.01127},
 primaryClass = {astro-ph.CO},
       adsurl = {https://ui.adsabs.harvard.edu/abs/2023MNRAS.522.2188S}
}

@article{Ghara_2024,
   title={Probing the intergalactic medium during the Epoch of Reionization using 21 cm signal power spectra},
   volume={687},
   ISSN={1432-0746},
   url={http://dx.doi.org/10.1051/0004-6361/202449444},
   DOI={10.1051/0004-6361/202449444},
   journal={A\&A},
   publisher={EDP Sciences},
   author={Ghara, R. and Shaw, A. K. and Zaroubi, S. and Ciardi, B. and Mellema, G. and Koopmans, L. V. E. and Acharya, A. and Choudhury, M. and Giri, S. K. and Iliev, I. T. and Ma, Q. and Mertens, F. G.},
   year={2024},
   month=jul, pages={A252} }

@ARTICLE{2022ApJ...924...51A,
       author = {{Abdurashidova}, Zara and {Aguirre}, James E. and {Alexander}, Paul and {Ali}, Zaki S. and {Balfour}, Yanga and {Barkana}, Rennan and {Beardsley}, Adam P. and {Bernardi}, Gianni and {Billings}, Tashalee S. and {Bowman}, Judd D. and {Bradley}, Richard F. and {Bull}, Philip and {Burba}, Jacob and {Carey}, Steve and {Carilli}, Chris L. and {Cheng}, Carina and {DeBoer}, David R. and {Dexter}, Matt and {de Lera Acedo}, Eloy and {Dillon}, Joshua S. and {Ely}, John and {Ewall-Wice}, Aaron and {Fagnoni}, Nicolas and {Fialkov}, Anastasia and {Fritz}, Randall and {Furlanetto}, Steven R. and {Gale-Sides}, Kingsley and {Glendenning}, Brian and {Gorthi}, Deepthi and {Greig}, Bradley and {Grobbelaar}, Jasper and {Halday}, Ziyaad and {Hazelton}, Bryna J. and {Heimersheim}, Stefan and {Hewitt}, Jacqueline N. and {Hickish}, Jack and {Jacobs}, Daniel C. and {Julius}, Austin and {Kern}, Nicholas S. and {Kerrigan}, Joshua and {Kittiwisit}, Piyanat and {Kohn}, Saul A. and {Kolopanis}, Matthew and {Lanman}, Adam and {La Plante}, Paul and {Lekalake}, Telalo and {Lewis}, David and {Liu}, Adrian and {Ma}, Yin-Zhe and {MacMahon}, David and {Malan}, Lourence and {Malgas}, Cresshim and {Maree}, Matthys and {Martinot}, Zachary E. and {Matsetela}, Eunice and {Mesinger}, Andrei and {Mirocha}, Jordan and {Molewa}, Mathakane and {Morales}, Miguel F. and {Mosiane}, Tshegofalang and {Mu{\~n}oz}, Julian B. and {Murray}, Steven G. and {Neben}, Abraham R. and {Nikolic}, Bojan and {Nunhokee}, Chuneeta D. and {Parsons}, Aaron R. and {Patra}, Nipanjana and {Pieterse}, Samantha and {Pober}, Jonathan C. and {Qin}, Yuxiang and {Razavi-Ghods}, Nima and {Reis}, Itamar and {Ringuette}, Jon and {Robnett}, James and {Rosie}, Kathryn and {Santos}, Mario G. and {Sikder}, Sudipta and {Sims}, Peter and {Smith}, Craig and {Syce}, Angelo and {Thyagarajan}, Nithyanandan and {Williams}, Peter K.~G. and {Zheng}, Haoxuan},
        title = "{HERA Phase I Limits on the Cosmic 21 cm Signal: Constraints on Astrophysics and Cosmology during the Epoch of Reionization}",
      journal = {\apj},
         year = 2022,
        month = jan,
       volume = {924},
       number = {2},
          eid = {51},
        pages = {51},
          doi = {10.3847/1538-4357/ac2ffc},
archivePrefix = {arXiv},
       eprint = {2108.07282},
 primaryClass = {astro-ph.CO},
       adsurl = {https://ui.adsabs.harvard.edu/abs/2022ApJ...924...51A}
}

@ARTICLE{munshi2024,
       author = {{Munshi}, S. and {Mertens}, F.~G. and {Koopmans}, L.~V.~E. and {Offringa}, A.~R. and {Semelin}, B. and {Aubert}, D. and {Barkana}, R. and {Bracco}, A. and {Brackenhoff}, S.~A. and {Cecconi}, B. and {Ceccotti}, E. and {Corbel}, S. and {Fialkov}, A. and {Gehlot}, B.~K. and {Ghara}, R. and {Girard}, J.~N. and {Grie{\ss}meier}, J.~M. and {H{\"o}fer}, C. and {Hothi}, I. and {M{\'e}riot}, R. and {Mevius}, M. and {Ocvirk}, P. and {Shaw}, A.~K. and {Theureau}, G. and {Yatawatta}, S. and {Zarka}, P. and {Zaroubi}, S.},
        title = "{First upper limits on the 21 cm signal power spectrum from cosmic dawn from one night of observations with NenuFAR}",
      journal = {\aap},
         year = 2024,
        month = jan,
       volume = {681},
          eid = {A62},
        pages = {A62},
          doi = {10.1051/0004-6361/202348329},
archivePrefix = {arXiv},
       eprint = {2311.05364},
 primaryClass = {astro-ph.CO},
       adsurl = {https://ui.adsabs.harvard.edu/abs/2024A&A...681A..62M}
}

@ARTICLE{2024MNRAS.530..191G,
       author = {{Ghara}, Raghunath and {Bag}, Satadru and {Zaroubi}, Saleem and {Majumdar}, Suman},
        title = "{The morphology of the redshifted 21-cm signal from the Cosmic Dawn}",
      journal = {\mnras},
         year = 2024,
        month = may,
       volume = {530},
       number = {1},
        pages = {191-202},
          doi = {10.1093/mnras/stae895},
archivePrefix = {arXiv},
       eprint = {2308.00548},
 primaryClass = {astro-ph.CO},
       adsurl = {https://ui.adsabs.harvard.edu/abs/2024MNRAS.530..191G}
}

@ARTICLE{2024MNRAS.534L..30A,
       author = {{Acharya}, Anshuman and {Mertens}, Florent and {Ciardi}, Benedetta and {Ghara}, Raghunath and {Koopmans}, L{\'e}on V.~E. and {Zaroubi}, Saleem},
        title = "{Revised LOFAR upper limits on the 21-cm signal power spectrum at z {\ensuremath{\approx}} 9.1 using machine learning and gaussian process regression}",
      journal = {\mnras},
         year = 2024,
        month = oct,
       volume = {534},
       number = {1},
        pages = {L30-L34},
          doi = {10.1093/mnrasl/slae078},
archivePrefix = {arXiv},
       eprint = {2408.10051},
 primaryClass = {astro-ph.CO},
       adsurl = {https://ui.adsabs.harvard.edu/abs/2024MNRAS.534L..30A}
}

@ARTICLE{emilio2024,
       author = {{Ceccotti}, E. and {Offringa}, A.~R. and {Koopmans}, L.~V.~E. and {Mertens}, F.~G. and {Mevius}, M. and {Acharya}, A. and {Brackenhoff}, S.~A. and {Ciardi}, B. and {Gehlot}, B.~K. and {Ghara}, R. and {Chege}, J.~K. and {Ghosh}, S. and {H{\"o}fer}, C. and {Hothi}, I. and {Iliev}, I.~T. and {McKean}, J.~P. and {Munshi}, S. and {Zaroubi}, S.},
        title = "{Spectral modelling of Cygnus A between 110 and 250 MHz: Impact on the LOFAR 21-cm signal power spectrum}",
      journal = {\aap},
         year = 2025,
        month = apr,
       volume = {696},
          eid = {A56},
        pages = {A56},
          doi = {10.1051/0004-6361/202453106},
archivePrefix = {arXiv},
       eprint = {2502.18459},
 primaryClass = {astro-ph.CO},
       adsurl = {https://ui.adsabs.harvard.edu/abs/2025A&A...696A..56C}
}

@ARTICLE{2025ApJ...988...84G,
       author = {{Gao}, Li-Yang and {Koopmans}, L{\'e}on V.~E. and {Mertens}, Florent G. and {Munshi}, Satyapan and {Li}, Yichao and {Brackenhoff}, Stefanie A. and {Ceccotti}, Emilio and {Chege}, J. Kariuki and {Acharya}, Anshuman and {Ghara}, Raghunath and {Giri}, Sambit K. and {Iliev}, Ilian T. and {Mellema}, Garrelt and {Zhang}, Xin},
        title = "{Extracting the Epoch of Reionization Signal with 3D U-Net Neural Networks Using a Data-driven Systematic Effect Model}",
      journal = {\apj},
         year = 2025,
        month = jul,
       volume = {988},
       number = {1},
          eid = {84},
        pages = {84},
          doi = {10.3847/1538-4357/ade2dc},
archivePrefix = {arXiv},
       eprint = {2412.16853},
 primaryClass = {astro-ph.IM},
       adsurl = {https://ui.adsabs.harvard.edu/abs/2025ApJ...988...84G}
}

@ARTICLE{2025A&A...699A.109G,
       author = {{Ghara}, R. and {Zaroubi}, S. and {Ciardi}, B. and {Mellema}, G. and {Giri}, S.~K. and {Mertens}, F.~G. and {Mevius}, M. and {Koopmans}, L.~V.~E. and {Iliev}, I.~T. and {Acharya}, A. and {Brackenhoff}, S.~A. and {Ceccotti}, E. and {Chege}, K. and {Georgiev}, I. and {Ghosh}, S. and {Hothi}, I. and {H{\"o}fer}, C. and {Ma}, Q. and {Munshi}, S. and {Offringa}, A.~R. and {Shaw}, A.~K. and {Pandey}, V.~N. and {Yatawatta}, S. and {Choudhury}, M.},
        title = "{Constraints on the state of the intergalactic medium at z{\ensuremath{\sim}}8 ‑ 10 using redshifted 21 cm observations with LOFAR}",
      journal = {\aap},
         year = 2025,
        month = jul,
       volume = {699},
          eid = {A109},
        pages = {A109},
          doi = {10.1051/0004-6361/202554163},
archivePrefix = {arXiv},
       eprint = {2505.00373},
 primaryClass = {astro-ph.CO},
       adsurl = {https://ui.adsabs.harvard.edu/abs/2025A&A...699A.109G}
}

@article{Furlanetto_2004,
doi = {10.1086/423025},
url = {https://dx.doi.org/10.1086/423025},
year = {2004},
month = {sep},
publisher = {},
volume = {613},
number = {1},
pages = {1},
author = {Furlanetto, Steven R. and Zaldarriaga, Matias and Hernquist, Lars},
title = {The Growth of H II Regions During Reionization},
journal = {The Astrophysical Journal}
}

@article{eilers2021detecting,
  title={Detecting and Characterizing Young Quasars. II. Four Quasars at z~ 6 with Lifetimes< 104 Yr},
  author={Eilers, Anna-Christina and Hennawi, Joseph F and Davies, Frederick B and Simcoe, Robert A},
  journal={The Astrophysical Journal},
  volume={917},
  number={1},
  pages={38},
  year={2021},
  publisher={IOP Publishing}
}

\end{document}